\documentclass[author,reqno]{nrc2} 

\usepackage{graphicx}
\usepackage{amssymb,amsmath}
\usepackage{cite}
\usepackage{url}
\usepackage{color}

\newcommand{\bea}{\begin{eqnarray}}
\newcommand{\eea}{\end{eqnarray}}
\newcommand{\be}{\begin{equation}}
\newcommand{\ee}{\end{equation}}
\newcommand{\rr}{\mathbf{r}}
\newcommand{\xx}{\mathbf{x}}
\newcommand{\sss}{\mathbf{s}}
\newcommand{\qq}{\mathbf{q}}
\newcommand{\kk}{\mathbf{k}}
\newcommand{\dd}{\mathrm{d}}
\newcommand{\ii}{\mathrm{i}}
\newcommand{\eee}{\mathrm{e}}

\DeclareMathOperator\asin{asin}

\DeclareMathOperator\argch{argch}
\DeclareMathOperator\argsh{argsh}
\DeclareMathOperator\ch{ch}
\DeclareMathOperator\sh{sh}
\DeclareMathOperator\re{Re}
\DeclareMathOperator\im{Im}

\begin{document}

\title{Bragg spectroscopy and pair-breaking-continuum mode in a superfluid Fermi gas}
\author{Yvan Castin} 
\address{Laboratoire Kastler Brossel, ENS-Universit\'e PSL, CNRS, Universit\'e de la Sorbonne et Coll\`ege de France, 24 rue Lhomond, 75231 Paris, France}
\shortauthor{Yvan Castin}

\begin{abstract}
The superfluid, pair condensed spin-$1/2$ Fermi gases are supposed to exhibit at nonzero wave vector a still unobserved collective excitation mode in their pair-breaking continuum. Using BCS theory at zero temperature and in the long wavelength limit, we predict that this mode is quantitatively observable (in frequency, width and spectral weight) in the response of a cold atom gas to a laser Bragg excitation, if one measures the perturbation induced on the order parameter modulus rather than on the density.
\end{abstract}

\maketitle

\section {Introduction}

We now know how to prepare in the laboratory a gas of fermionic cold atoms of spin $ 1/2 $ trapped in a flat-bottom box potential \cite{buxida1}, thus spatially homogeneous \cite{buxida2, buxida3, buxida4}. These atoms are subjected to an attractive interaction in the $ s $ wave between opposite spin states $ \uparrow $ and $ \downarrow $ of van der Waals type, of negligible range $b$ and of scattering length $ a $ tunable with a magnetic Feshbach resonance \cite{Thomas2002, Salomon2003, Grimm2004b, Ketterle2004, Salomon2010, Zwierlein2012}. One can arrange for the gas to be unpolarized, that is, it has the same number of particles in $ \uparrow $ and $ \downarrow $. At the very low temperatures experimentally attained, we can then assume, as a first approximation, that all fermions assemble in $ \uparrow \downarrow $ bound pairs, equivalent for our neutral system to  superconductor Cooper pairs, these pairs forming furthermore a condensate and a superfluid, as predicted by BCS theory.

At the thermodynamic limit, with the excitation vector $ \qq $ fixed, the excitation spectrum of the zero temperature system has a pair-breaking continuum of the form $ \varepsilon _{\qq / 2 + \kk } + \varepsilon _{\qq / 2- \kk} $, where $ k \mapsto \varepsilon_\kk $ is the dispersion relation of a broken pair fragment and the relative wave vector $ \kk $ of the two fragments spans the entire three-dimensional Fourier space.  We limit ourselves here to the usual case, where the bound pairs are of sufficiently large radius (with respect to the average distance between fermi\-ons) to exhibit a clear composite boson character, i.e. $ k \mapsto \varepsilon_ \kk $ reaches its minimum in a wave number $ k_0> 0 $. If $ 0 <q <2 k_0 $, the density of states of the pair-breaking continuum then exhibits, on the real energy axis, two singularity points $ \varepsilon_1 (q) <\varepsilon_2 (q) $, and even a third one $ \varepsilon_3 (q)> \varepsilon_2 (q) $ for $ q / k_0 $ sufficiently small. By analytic continuation of the eigenvalue equation through the interval $ [\varepsilon_1 (q), \varepsilon_2 (q)] $, we find that this continuum contains a pair-breaking collective excitation mode of complex energy $ z_\qq $ deviating quadratically with $ q $ from its $ 2 \Delta $ limit at $ q = 0 $, where $ \Delta $ is the order parameter of the pair condensate, taken real positive at equilibrium. This is predicted both in the weak coupling limit $ \Delta \ll \varepsilon_F $ \cite{AndrianovPopov}, where $ \varepsilon_F $ is the Fermi energy of the gas, and in the strong coupling limit $ \Delta \approx \varepsilon_F $ \cite{PRL2019, CRAS2019}; according to the used time-dependent BCS theory, it suffices that the chemical potential of the gas is positive, $ \mu> 0 $, so that $k_0>0$. Note that the continuum mode is often called the amplitude mode, or even the Higgs mode \cite{Varma}, but the analogy with high energy physics is only approximate \cite{Benfatto} and the dispersion relation given in reference \cite{Varma} is incorrect. Note also that other continuum modes can be obtained by analytic continuation through the intervals $ [\varepsilon_2 (q), + \infty [$, $ [\varepsilon_2 (q), \varepsilon_3 (q)] $ and $ [\varepsilon_3 (q), + \infty [$ (if $ \varepsilon_3 (q) $ exists), and that the regime $ k_0 = 0 $ ($ \mu <0 $ according to BCS theory) also exhibits continuum modes by continuation through $ [\varepsilon_1 (q), + \infty [$ (in this case there is only one singularity point), even within the limit $ \mu / \varepsilon_F \to - \infty $ where the bound pairs reduce to elementary bosons \cite {CRAS2019}. These exotic modes generally have a complex energy $ z_ \qq $ of real part far from their analytic continuation interval at low $ q $ and of high imaginary part (in absolute value) at large $ q $, which makes them difficult to observe according to the criteria of reference \cite {PRL2019}; so we ignore them here.

The question is how to get evidence for the ordinary continuum branch. The question is important because the branch is currently unobserved: the oscillations of the order parameter at angular frequency $ 2 \Delta / \hbar $ detected in a superconductor \cite{Shimano, Sacuto} or in a cold fermionic atom gas \cite{Koehl} after a spatially homogeneous excitation have a time dependence damped as a power law $ \sin (2 \Delta t / \hbar + \phi) / t^\alpha $ rather than purely sinusoidal \cite{Kogan, Altshuler, Gurarie}, and do not result from a discrete mode of the superfluid but simply from a generic effect of the edge $ 2 \Delta $ of the pair-breaking continuum \cite{Orbach, Stringari}; the key point at $ q = 0 $ is that, as the theory predicts, the eigenenergy equation has the zero energy as only root (this is the starting point of the Anderson-Bogolioubov acoustic branch), in particular the energy $ 2 \Delta $ is not a root, even after analytic continuation to the lower half-plane \cite{PRL2019}.  Under certain conditions however, the $ \chi $ linear response (or susceptibility) functions of the system to an excitation of angular frequency  $ \omega $ and nonzero wavevector $\qq$ must exhibit, as functions of the angular frequency, a peak centered near $ \omega = \re z_\qq / \hbar $ and of approximate half-width $ \im z_\qq / \hbar $, above the broad continuum response background, which is characteristic of a modal contribution. This is the case for the mod\-ulus-modulus response function $ \chi_{| \Delta | | \Delta |} (\qq, \omega) $, where we look at the effect on the modulus of the order parameter of a modulus excitation of the order parameter {\sl via} a spatial and temporal modulation of the scattering length \cite{PRL2019}; such excitation is however difficult to implement. On the other hand, the density excitation of a gas of cold atoms by a Bragg pulse, by means of two laser beams of angular frequency difference $ \omega $ and of wave vector difference $ \qq $, is a technique well established in the laboratory, which gave rise to a real Bragg spectroscopy \cite{Bragg1, Bragg2, Bragg3, Bragg4, Bragg5}. Depending on whether we measure the variation of the total density $ \rho $ of the gas (by absorption or dispersion of a laser beam \cite{KetterleVarenna}) or the modulus $ | \Delta | $ of its order parameter (by interferometry \cite{Iacopo} or bosonization of bound pairs $ \uparrow \downarrow $ by fast Feshbach ramp \cite{KetterleVarenna,Ketterletourb}) as a result of the Bragg pulse, we access the response function $ \chi_{ \rho \rho} (\qq, \omega) $ or $ \chi_{| \Delta | \rho} (\qq, \omega) $. On the one hand, the density-density susceptibility of a superfluid fermion gas has been the subject of numerous theoretical \cite{crrth1, crrth2, crrth3, crrth4, crrth5, crrth6,crrth7} and experimental \cite{Bragg3, Bragg4, Bragg5} studies, but without the slightest attention being paid to the continuum mode; on the other hand, the modulus-density susceptibility has rarely been calculated, and to our knowledge never measured with cold atoms. The purpose of this paper is to fill these two gaps, at least theoretically. 

\section {Response Functions in BCS Theory} 

Our pairwise condensed fermion gas, initially prepared at equilibrium at zero temperature, is subjected to density excitation, that is, to a perturbation of its Hamiltonian of the form 
\be
\label{eq:001}
\hat{W}(t)=\int \dd^3r\, U(\rr,t) \sum_{\sigma=\uparrow,\downarrow} \hat{\psi}_\sigma^{\dagger}(\rr) \hat{\psi}_\sigma(\rr)
\ee
where the real potential $ U (\rr, t) $ depends on time and space, and the $ \hat {\psi} _\sigma (\rr) $ and $ \hat {\psi}^\dagger_\sigma (\rr) $ field operators, written in Schr\"odinger's picture, annihilate and create a fermion in spin state $ \sigma $ at point $ \rr $ and obey the usual fermionic anticommutation relations. When $ U (\rr, t) $ is weak enough or applied for a short enough time, the system's response to an observable $ \hat {O} $ is linear, i.e.\, the deviation $ \delta \langle \hat {O} \rangle $ of the average value of $ \hat {O} $ from its equilibrium value is a linear functional of $ U $, described by a susceptibility $ \chi_{ O \rho} $. We limit ourselves here to two observables, the total density $ \rho $ and the modulus $ | \Delta | $ of the complex order parameter $ \Delta $ defined in reference \cite{RMP}: 
\bea
\label{eq:002a}
\delta\rho(\rr,t) &\!\!\!=\!\!\!& \int\dd^3r'\int \dd t' \chi_{\rho\rho}(\rr-\rr',t-t') U(\rr',t') \\
\label{eq:002b}
\delta|\Delta|(\rr,t) &\!\!\!=\!\!\!& \int\dd^3r'\int \dd t' \chi_{|\Delta|\rho}(\rr-\rr',t-t') U(\rr',t') 
\eea
As the initial state of the system is stationary and spatially homogeneous, the susceptibilities depend only on the difference of time and position; they are also causal and therefore retarded (zero if $ t <t '$). In practice, the Bragg excitation mentioned in the introduction corresponds to the lightshift potential $ U (\rr, t) = U_0 \eee^{\ii (\qq \cdot \rr- \omega t)} + \mbox {c.c.} $, where the amplitude $ U_0 $ is complex. It gives access, as shown by the insertion of $ U (\rr, t) $ in (\ref {eq:002a}) and (\ref {eq:002b}), to the spatiotemporal Fourier transform of susceptibilities: 
\be
\label{eq:003}
\chi(\qq,\omega) \equiv \int\dd^3r\int\dd t\, \eee^{\ii[(\omega+\ii\eta)t-\qq\cdot\rr]} \chi(\rr,t) \quad\quad(\eta\to 0^+)
\ee
The factor $ \eee^{- \eta t} $ ensuring the convergence of the integral over time is usual for retarded Green functions \cite{CCT}. 

To obtain an approximate expression of susceptibilities using time-dependent BCS variational theory, it is convenient to use a cubic lattice model of spacing $ b $ in the quantization volume $ [0, L]^3 $ with periodic boundary conditions, by making $ b $ tend to zero and $ L $ tend to infinity at the end of the calculations. The fermions of mass $ m $ have the free space dispersion relation $ \kk \mapsto E_\kk = \hbar^2 k^2 / 2m $ on the first Brillouin zone $ \mathcal {D} = [ - \pi / b, \pi / b [^ 3 $, and it is extended by periodicity beyond. They interact by the contact binary potential $ V (\rr_i, \rr_j) = g_0 \delta _{\rr_i, \rr_j} / b^3 $, with a bare coupling constant $ g_0 $ adjusted to reproduce the scattering length $ a $ of the experiment \cite{Houches, livreZwerger}: $ 1 / g_0 = 1 / g- \int _{\mathcal {D}} \frac {\dd^3k} {(2 \pi)^3} \frac {1} {2E_\kk} $ where $ g = 4 \pi \hbar^2a / m $ is the effective coupling constant. The grand canonical ground state of the gas of chemical potential $ \mu $ (of arbitrary sign in this section) is approximated by the usual state $ | \psi_0 \rangle $, a coherent state of fermion pairs breaking $U(1)$ symmetry: this is the vacuum of the  fermionic quasi-particle annihilation operators $ \hat {\gamma} _{\kk \sigma} $ defined below. The BCS variational ansatz extends to the time-dependent case \cite{Varenna}, and the order parameter is simply 
\be
\Delta(\rr,t)=g_0 \langle\psi(t)|\hat{\psi}_\downarrow(\rr)\hat{\psi}_\uparrow(\rr)|\psi(t)\rangle
\ee
To obtain $ \chi (\qq, \omega) $, the simplest is to consider a percussive excitation in time with a well-defined nonzero wave vector, $ U (\rr, t) = \hbar \epsilon \cos (\qq \cdot \rr) \delta (t) $, with $ \epsilon \to 0 $. The time-dependent perturbation theory gives the state vector just after the perturbation to first order in $ \epsilon $: 
\begin{multline}
|\psi(0^+)\rangle\simeq\left[1\!-\!\ii\epsilon\!\int\!\dd^3r\, \cos(\qq\cdot\rr) \sum_{\sigma} \hat{\psi}_\sigma^{\dagger}(\rr) \hat{\psi}_\sigma(\rr)\right]|\psi(0^-)\rangle \\
\hspace{-3mm}\simeq \left[1\!+\!\frac{\ii\epsilon}{2}\sum_{\kk} (U_+V_-\!+\!U_-V_+) (\hat{\gamma}^\dagger_{+\uparrow}\hat{\gamma}^\dagger_{-\downarrow}\!+\!\qq\leftrightarrow\!-\qq) \right]|\psi_0\rangle
\label{eq:005}
\end{multline}
Here, the indices $ + $ and $ - $ refer to the wave numbers $ \qq / 2 + \kk $ and $ \qq / 2- \kk $, the coefficients $ U_\kk = [\frac {1} {2} (1+ \xi_\kk / \varepsilon_\kk)]^{ 1/2} $ and $ V_\kk = [\frac {1} {2} (1- \xi_\kk / \varepsilon_\kk)]^{1/2} $ are the amplitudes of the quasi-particle modes on particles and holes, and $ \kk \mapsto \varepsilon_\kk = (\xi_\kk^2 + \Delta^2)^{1/2} $ is their BCS dispersion relation, with $ \xi_\kk = E_\kk- \mu + g_0 \rho / 2 $.\footnote {It was necessary to use the modal expansions of the field operators, $ \hat {\psi} _\uparrow (\rr) = L^{- 3/2} \sum_\kk (\hat {\gamma} _{\kk \uparrow} U_\kk- \hat {\gamma}^\dagger _{- \kk \downarrow} V_\kk) \eee^{ \ii \kk \cdot \rr} $ and $ \hat {\psi} _\downarrow (\rr) = L^{- 3/2} \sum_\kk (\hat {\gamma} _{\kk \downarrow} U_\kk + \hat {\gamma}^\dagger _{- \kk \uparrow} V_\kk) \eee^{\ii \kk \cdot \rr} $.} The evolution of density and order parameter for a very weak coherent state of {\sl quasiparticle} pairs as (\ref {eq:005}) (which remains of course a strong coherent state of fermionic-atom pairs) was calculated with time-dependent BCS theory \cite{HadrienThese, Annalen}; by particularizing the general expressions of  reference \cite{CRAS2019}, we find for $ t> 0 $: 
\be
\begin{pmatrix}
2\ii\Delta(\delta\theta)_\qq(t) \\
2(\delta|\Delta|)_\qq(t) \\
(\delta\rho)_\qq(t)
\end{pmatrix}
= (-\ii\epsilon) \int_{\ii\eta+\infty}^{\ii\eta-\infty} \frac{\dd z}{2\ii\pi} \frac{\eee^{-\ii z t/\hbar}}{M(z,\qq)} 
\begin{pmatrix}
\Sigma_{13}(z,\qq) \\
\Sigma_{23}(z,\qq) \\
\Sigma_{33}(z,\qq) 
\end{pmatrix}
\ee
where $ \theta (\rr, t) = \arg \Delta (\rr, t) $ is the phase of the order parameter and $ X_\qq $ is the Fourier coefficient of $ X (\rr) $ on the plane wave $ \eee^{\ii \qq \cdot \rr} $. We had to introduce the $ 3 \times 3 $ matrix, function of the complex energy $ z $ in the upper half-plane and of the wave number, 
\be
\label{eq:007}
M(z,\qq) = 
\begin{pmatrix}
\Sigma_{11}(z,\qq) & \Sigma_{12}(z,\qq) & \phantom{1}-g_0 \Sigma_{13}(z,\qq) \\
\Sigma_{12}(z,\qq) & \Sigma_{22}(z,\qq) & \phantom{1}-g_0 \Sigma_{23}(z,\qq) \\
\Sigma_{13}(z,\qq) & \Sigma_{23}(z,\qq) & 1-g_0\Sigma_{33}(z,\qq) 
\end{pmatrix}
\ee
described by the six independent coefficients: 
\begin{equation}
\!\!\!\!\begin{array}{l}
\displaystyle
\Sigma_{11}(z,\qq)=\!\!\int_{\mathcal{D}}\!\frac{\dd^3k}{(2\pi)^3} \frac{(\varepsilon_++\varepsilon_-)(\varepsilon_+\varepsilon_-+\xi_+\xi_-+\Delta^2)}{2\varepsilon_+\varepsilon_-[z^2-(\varepsilon_++\varepsilon_-)^2]}+\frac{1}{2\varepsilon_\kk} \\
\displaystyle
\Sigma_{22}(z,\qq)=\!\!\int_{\mathcal{D}}\!\frac{\dd^3k}{(2\pi)^3} \frac{(\varepsilon_++\varepsilon_-)(\varepsilon_+\varepsilon_-+\xi_+\xi_--\Delta^2)}{2\varepsilon_+\varepsilon_-[z^2-(\varepsilon_++\varepsilon_-)^2]}+\frac{1}{2\varepsilon_\kk}\\
\displaystyle
\Sigma_{33}(z,\qq)=\!\!\int_{\mathcal{D}}\!\frac{\dd^3k}{(2\pi)^3} \frac{(\varepsilon_++\varepsilon_-)(\varepsilon_+\varepsilon_--\xi_+\xi_-+\Delta^2)}{2\varepsilon_+\varepsilon_-[z^2-(\varepsilon_++\varepsilon_-)^2]}\\
\displaystyle
\Sigma_{12}(z,\qq)=\!\!\int_{\mathcal{D}}\!\frac{\dd^3k}{(2\pi)^3} \frac{z(\xi_+\varepsilon_-+\xi_-\varepsilon_+)}{2\varepsilon_+\varepsilon_-[z^2-(\varepsilon_++\varepsilon_-)^2]}\\
\displaystyle
\Sigma_{13}(z,\qq)=\!\!\int_{\mathcal{D}}\!\frac{\dd^3k}{(2\pi)^3} \frac{z \Delta (\varepsilon_++\varepsilon_-)}{2\varepsilon_+\varepsilon_-[z^2-(\varepsilon_++\varepsilon_-)^2]}\\
\displaystyle
\Sigma_{23}(z,\qq)=\!\!\int_{\mathcal{D}}\!\frac{\dd^3k}{(2\pi)^3} \frac{\Delta (\varepsilon_++\varepsilon_-)(\xi_++\xi_-)}{2\varepsilon_+\varepsilon_-[z^2-(\varepsilon_++\varepsilon_-)^2]}
\end{array}
\end{equation}
Specializing (\ref {eq:002a}) and (\ref {eq:002b}) to the percussive excitation considered, it comes easily for our lattice model: 
\bea
\label{eq:009a}
\chi_{\rho\rho}(\qq,\omega) = (0,0,1) \cdot \frac{2}{M(z,\qq)}
\left.
\begin{pmatrix}
\Sigma_{13}(z,\qq) \\
\Sigma_{23}(z,\qq) \\
\Sigma_{33}(z,\qq) 
\end{pmatrix}\right\vert_{z=\hbar\omega+\ii\eta} \\
\label{eq:009b}
\chi_{|\Delta|\rho}(\qq,\omega) = (0,1,0)\cdot \frac{1}{M(z,\qq)} 
\left.
\begin{pmatrix}
\Sigma_{13}(z,\qq) \\
\Sigma_{23}(z,\qq) \\
\Sigma_{33}(z,\qq)
\end{pmatrix}\right\vert_{z=\hbar\omega+\ii\eta}
\eea
This we will use for a continuous space in what follows. 

\section {In the BEC-BCS crossover} 
\label {sec: CBE-BCS} 

In the one-parameter space measuring the interaction strength, what we call the BEC-BCS crossover is the intermediate zone between the strong attraction limit $ k_F a \to 0^+ $, where the ground state of the system is a Bose-Einstein condensate (BEC) of $ \uparrow \downarrow $ dimers of size $a$ small compared to the mean distance between particles, and the limit of weak attraction $ k_F a \to 0^- $, where the ground state is a BCS state of bound pairs $ \uparrow \downarrow $ of size $ \xi \approx \hbar^2 k_F / m \Delta $ much larger than the interatomic distance. Here $ k_F = (3 \pi^2 \rho)^{1/3} $ is the Fermi wave number of the gas. The crossover thus corresponds to the regime $ 1 \lesssim k_F | a | $, which is also the one in which the superfluid fermionic cold atomic gases are prepared in practice, to avoid high losses of particles by three-body collision in the BEC limit and too low critical temperatures in the BCS limit. 

However, our lattice model must always have a lattice spacing $ b \ll 1 / k_F $ to reproduce the physics of continuous space. So we also have $ b \ll | a | $, and we are led to make $ b $ tend to zero at fixed scattering length. We then replace the first Brillouin zone $ \mathcal {D} $ with the whole Fourier space $ \mathbb {R}^3 $. In the definition of $ \Sigma_{ij} $, this does not lead to any ultraviolet divergence; this triggers one in the expression of $ 1 / g_0 $, which makes $ g_0 $ tend to zero in matrix (\ref {eq:007}): 
\be
\label{eq:010}
g_0\to 0
\ee
The dispersion relation of BCS excitations is reduced to $ \varepsilon_k = [(E_k- \mu)^2 + \Delta^2]^{1/2} $; it has a minimum $ \Delta $ at a wave number $ k_0> 0 $, and the gas has a continuum collective excitation branch starting at $ 2 \Delta $ \cite{PRL2019, CRAS2019}, when the chemical potential $ \mu $ is $> 0 $, which we will assume now. Similarly, expressions (\ref {eq:009a}) and (\ref {eq:009b}) of susceptibilities are simplified as follows: 
\be
\label{eq:011}
\chi_{\rho\rho}= \frac{2\left|\begin{array}{lll} \Sigma_{11} & \Sigma_{12} & \Sigma_{13}\\ \Sigma_{12} & \Sigma_{22} & \Sigma_{23} \\ \Sigma_{13} & \Sigma_{23} & \Sigma_{33} \end{array}\right|}{\left|\begin{array}{ll} \Sigma_{11} & \Sigma_{12}\\ \Sigma_{12} & \Sigma_{22} \end{array}\right|}\,, \quad
\chi_{|\Delta|\rho}=  \frac{\left|\begin{array}{ll} \Sigma_{11} & \Sigma_{13}\\ \Sigma_{12} & \Sigma_{23} \end{array}\right|}
{\left|\begin{array}{ll} \Sigma_{11} & \Sigma_{12}\\ \Sigma_{12} & \Sigma_{22} \end{array}\right|}
\ee
where $ | A | $ is the determinant of matrix $ A $, and where we have made implicit the dependence of $ \chi $ on $ (\qq, \omega) $ and $ \Sigma_{ij} $ on $ (z, \qq) $ to shorten the notations.\footnote {\label {note:Cramer} Indeed, the vector $ \xx = M^{- 1} \sss $ is the solution of the system $ M \xx = \sss $, which one solves by the method of Cramer, with $ \sss $ the column vector of (\ref {eq:009a}) and (\ref {eq:009b}), to obtain its coordinates $ x_2 $ and $ x_3 $.} The value of $\chi_{\rho\rho}$ given agrees with equation (118) of reference \cite{crrth7}.

\begin {figure} [t] 
\centerline {\includegraphics [width = 4.25cm, clip =] {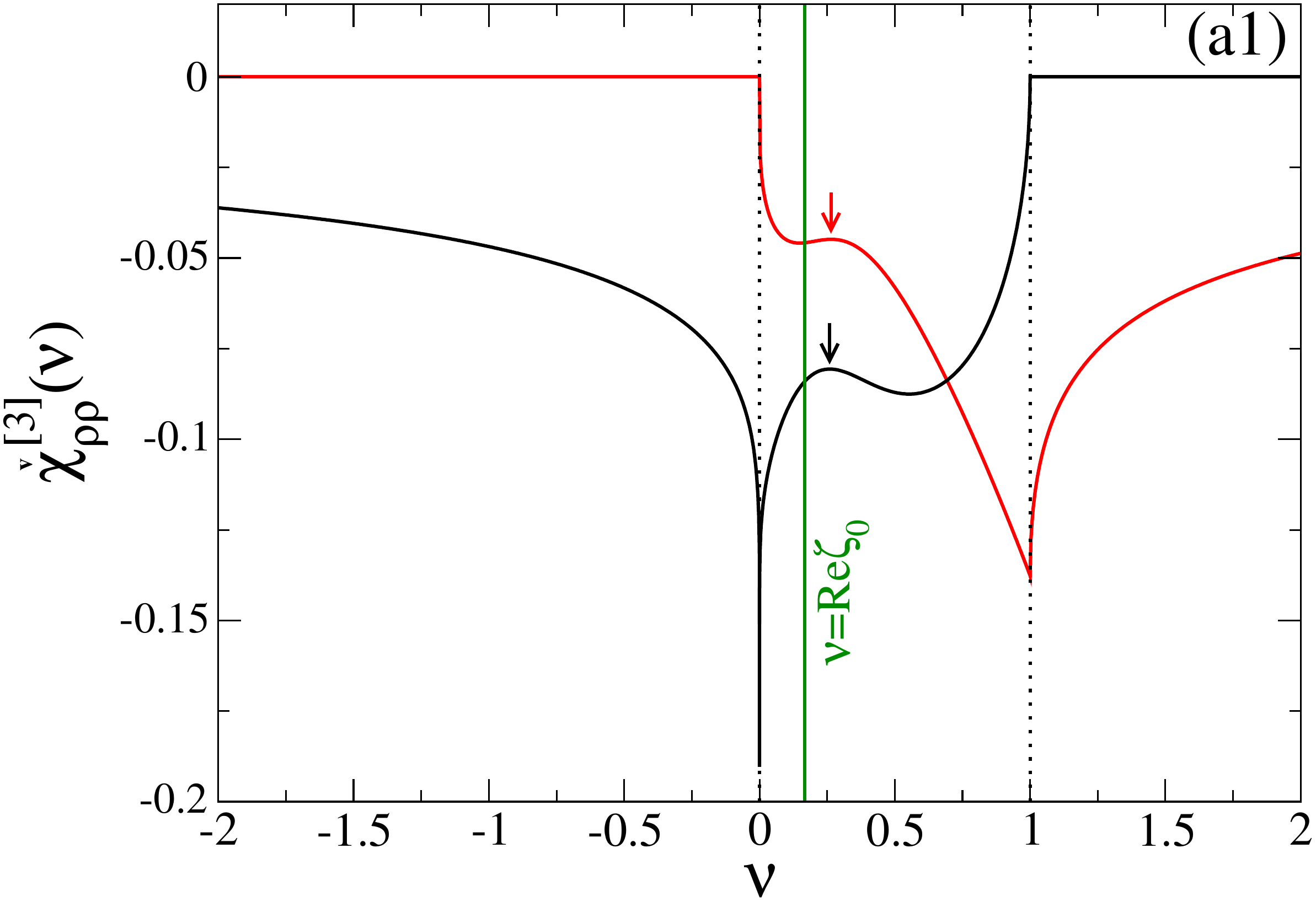} \hspace {3mm} \includegraphics [ width = 4.25cm, clip =] {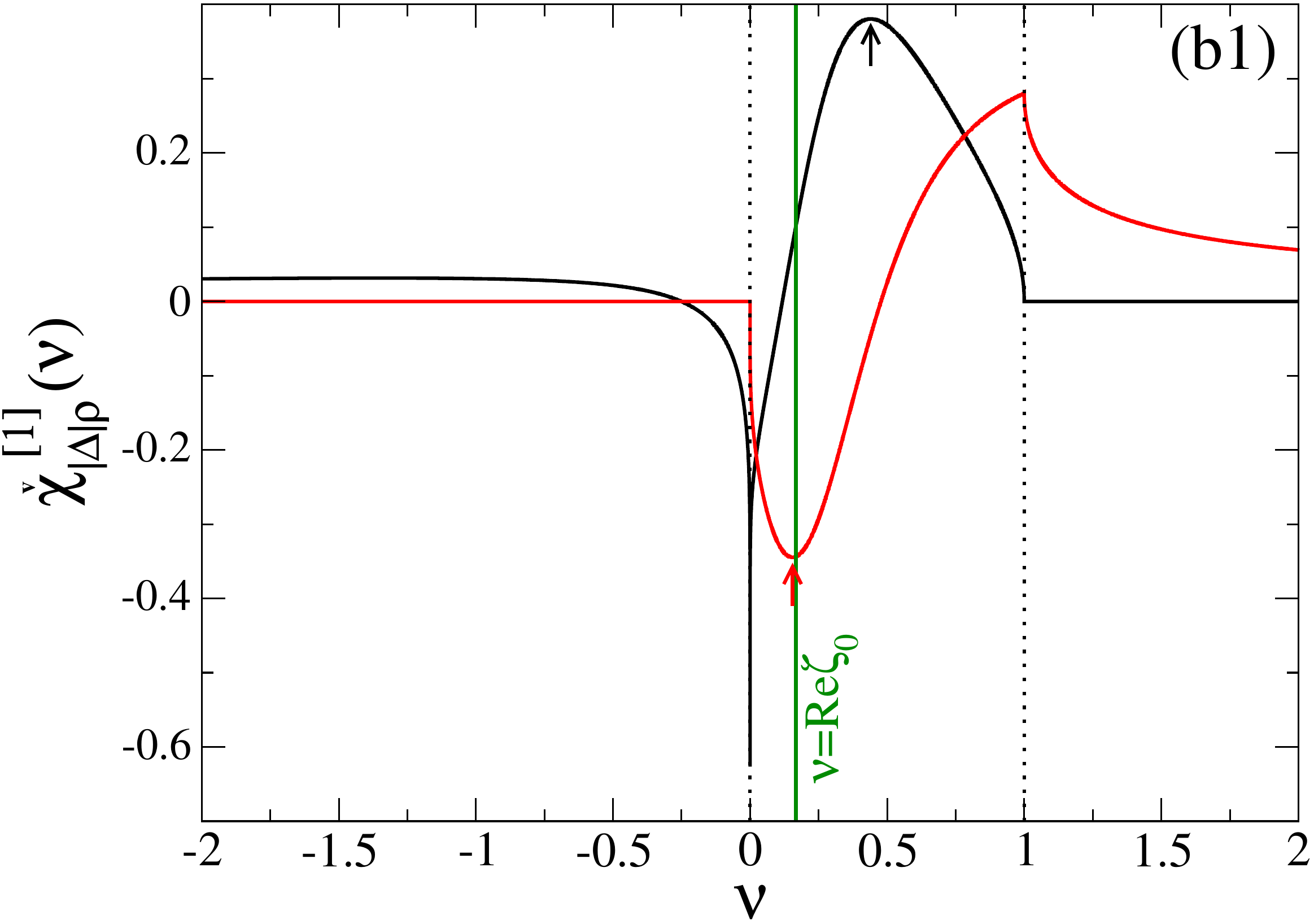}} 
\centerline {\includegraphics [width = 4.25cm, clip =] {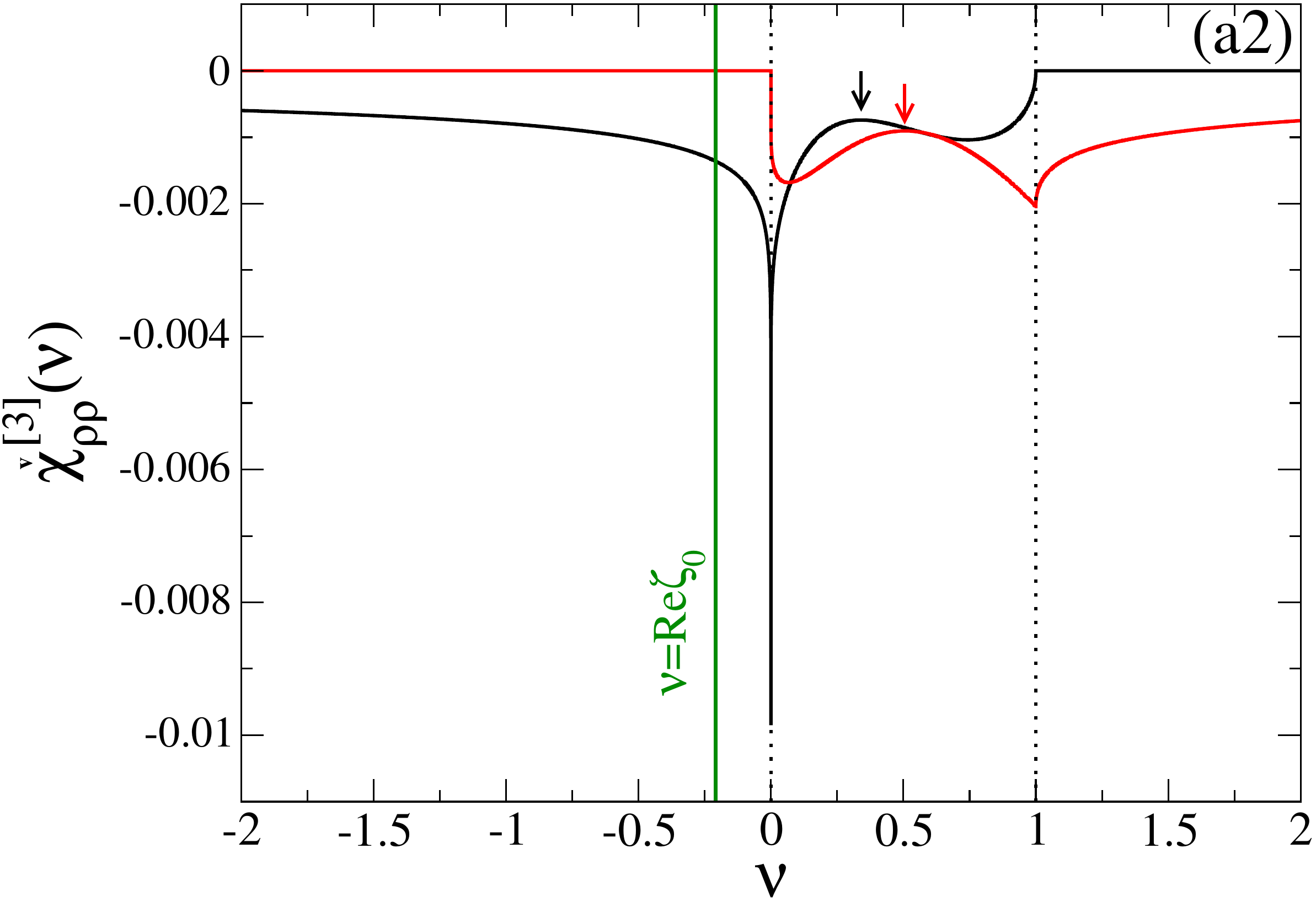} \hspace {3mm} \includegraphics [width = 4.25cm, clip =] {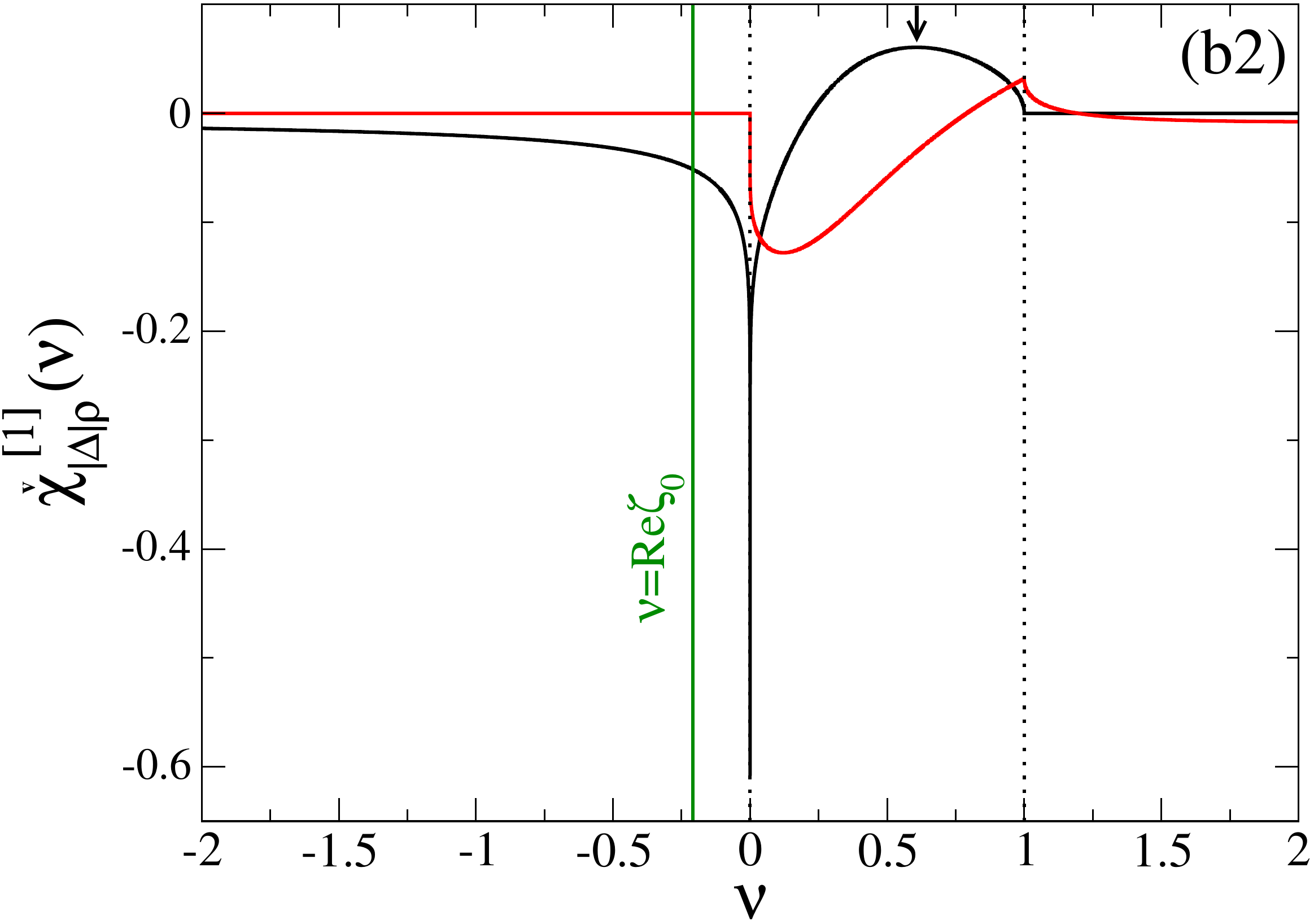}} 
\caption {First coefficient sensitive to the continuum mode in the low-wavenumber expansion $ q $ (\ref {eq:018a}, \ref {eq:018b}) of the density-density (column a, order $ q^3 $) and modulus-density (column b, order $ q $) response functions, on the $ \mu> 0 $ side of the BEC-BCS crossover (\ref {eq:010}), for $ \check {\Delta} = 1/2 $ (line 1) and $ \check {\Delta} = 2 $ (line 2), as a function of the reduced frequency $ \nu $ of equation (\ref {eq:013}) (using the variable $ \nu $ rather than $ \omega $ amounts to looking cleverly at the angular-frequency axis around $ 2 \Delta / \hbar $ with divergent magnification $ \propto q^{-2} $ exactly compensating for the narrowing of the continuum mode when $q \to 0$).  Black full line: real part; red full line: imaginary part. Vertical dots: positions $ \nu = 0 $ and $ \nu = 1 $ of the singularities. Green vertical line: reduced real part $ \re \zeta_0 $ of the complex energy of the continuum mode. The extrema indicated by an arrow are a physical mark of the continuum mode on the real frequency axis (see text and figure \ref {fig:lie}). Czech accent: $ \Delta $ rescaled by $ \mu $, $ q $ rescaled by $ k_0 = (2m \mu)^{1/2} / \hbar $, $ \chi _{\rho \rho} (\qq, \omega) $ rescaled by $ k_0^3 / \mu $; $ \chi_{| \Delta | \rho} (\qq, \omega) $ is already dimensionless.} 
\label {fig:gra} 
\end {figure}

Let's look for the possible signature of the continuum mode in the low-wavenumber response functions, $ q \to 0 $, where the imaginary part of the complex energy $ z_\qq $ is the lowest. This is where the mode has {\sl a priori} the best chance of emerging from the broad response background of the continuum as a narrow peak in the $ \omega $ domain. In this limit, under the condition $ q \ll k_0 \min (\Delta / \mu, (\mu / \Delta)^{1/2}) $ \cite{CRAS2019}, the continuum branch has a quadratic dispersion in $ q $: 
\be
\label{eq:012}
z_\qq \underset{q\to 0}{=} 2\Delta + \zeta_0 \frac{\hbar^2 q^2}{2m} \frac{\mu}{\Delta} + O(q^3)
\ee
The coefficient $ \zeta_0 $ is a solution in the lower complex half-plane of a transcendental equation given in reference \cite{PRL2019} (generalizing that of \cite{AndrianovPopov} to the BEC-BCS crossover). It is plotted as a function of the interaction strength in figure \ref{fig:fit}, and its limiting behaviors are given in \cite{PRL2019}. We will remember here that its real part is positive for $ \Delta / \mu <$ 1.21, and negative otherwise. Let's calculate the response functions on the real axis of angular frequencies near the continuum mode, by imposing the same wave number scale law as in (\ref {eq:012}): 
\be
\label{eq:013}
\hbar\omega \equiv 2\Delta + \nu \frac{\hbar^2q^2}{2m} \frac{\mu}{\Delta}\quad\quad (\nu\in\mathbb{R})
\ee
that is, by making $ q $ tend to zero at an arbitrary fixed reduced frequency $ \nu $.  This amounts to observing the angular-frequency axis around $ 2 \Delta / \hbar $ under a diverging magnification $ \propto q^{-2}$. In the following, it will be convenient to set 
\be
\label{eq:014}
\zeta=\nu+\ii\eta \quad\quad(\eta\to 0^+)
\ee
by analogy with $ z = \hbar \omega + \ii \eta $ in (\ref {eq:009a}) and (\ref {eq:009b}).  Since the singularity points $ \varepsilon_1 (q) $ and $ \varepsilon_2 (q) $ of the continuum density of states at fixed $ \qq $ mentioned in the introduction verify $ \varepsilon_1 (q) = 2 \Delta $ and $ \varepsilon_2 (q) = 2 \Delta + (\mu / \Delta) \hbar^2q^2/2m + O (q^4) $ \cite{PRL2019}, we expect in the limit $ q\to 0 $ that the response functions exhibit frequency singularities in $ \nu = 0 $ and $ \nu = 1 $, and that it is necessary to carry out the analytic continuation of the eigenvalue equation to $ \im \zeta <0 $ passing between the points $ \nu = 0 $ (i.e. $ \hbar \omega = \varepsilon_1 (q) $) and $ \nu = 1 $ (i.e. $ \hbar \omega = \varepsilon_2 (q) + O (q^4) $) to find the continuum mode. The fact that these singularities remain at finite $ \nu $ when $ q \to 0 $ makes it possible to immediately eliminate a false hope: even if its energy width tends towards zero like $ q^2 $, the continuum mode cannot lead to a very narrow peak in relative value in the frequency response functions, because the ``broad" contribution of the continuum has a structure varying at the same scale $ \propto q^2 $. On the other hand, the third singularity point $ \varepsilon_3 (q) = (\mu^2 + \Delta^2)^{1/2} + O(q^2) $ is irrelevant because rejected at $ \nu = + \infty $ by the change of scale (\ref{eq:013}).  For the calculation of $ \chi $ itself, the method for expanding the quantities $ \Sigma_{ij} $ in powers of $ q $ at fixed  $ \nu $ is known \cite{PRL2019}: it is not enough to naively expand the integrals over $ \kk $ under the integral sign in powers of $ q $, but it is necessary to treat separately the $\kk$-wave vector shell of thickness $ \propto q $ around the sphere $ k = k_0 $, which gives in general the dominant contribution because the energy denominators take extremely low values there, of the order of $ q^2 $.\footnote {After passing in spherical coordinates of axis $ \qq $, one separates the domain of integration on the modulus $ k $ in the two components $ I = [k_0-Aq, k_0 + Aq] $ and $ J = \mathbb {R}^+ \setminus I $, where $ A \gg 1 $ is fixed. On $ J $, we expand directly the integral in powers of $ q $ at fixed $ k $. On $ I $, we make the change of variable $ k = k_0 + qK $ then we expand the integral in powers of $ q $ at fixed $ K $. We collect the contributions of $ I $ and $ J $ order by order in $ q $, then we make $ A $ tend to $ + \infty $ in the coefficients of the monomials $ q^n $. In results (\ref {eq:017}), the contribution of $ J $ is negligible except in $ \delta \Sigma_{23}^{[2]}$ and $ \delta \Sigma_{33}^{[ 2,3]} $.} The forms (\ref {eq:011}) would lead to rather long calculations because, at the dominant order in $ q $, the first and the last column of determinants at the numerators are equivalent ($ \Sigma_{i3} \sim \Sigma_{1i} $, $ 1 \leq i \leq 3 $), which gives a zero result and forces us to get the subdominant orders of $ \Sigma_{ij} $. We can fortunately perform clever linear combinations without changing the value of these determinants, subtracting the first column from the last one and then, only in $ \chi _{\rho \rho} $, subtracting the first line from the third one: 
\be
\label{eq:015}
\chi_{\rho\rho}\!\!=\!\frac{2\left|\begin{array}{lll} \Sigma_{11} & \Sigma_{12} & \delta\Sigma_{13}\\ \Sigma_{12} & \Sigma_{22} & \delta\Sigma_{23} \\ \delta\Sigma_{13} & \delta\Sigma_{23} & \delta\Sigma_{33} \end{array}\right|}{\left|\begin{array}{ll} \Sigma_{11} & \Sigma_{12}\\ \Sigma_{12} & \Sigma_{22} \end{array}\right|}, 
\chi_{|\Delta|\rho}\!\!=\!\frac{\left|\begin{array}{ll} \Sigma_{11} & \delta\Sigma_{13}\\ \Sigma_{12} & \delta\Sigma_{23} \end{array}\right|}
{\left|\begin{array}{ll} \Sigma_{11} & \Sigma_{12}\\ \Sigma_{12} & \Sigma_{22} \end{array}\right|}\hspace{-3mm}
\ee
with 
\be
\label{eq:016}
\delta\Sigma_{13}\!\equiv\! \Sigma_{13}-\Sigma_{11}, 
\delta\Sigma_{23}\!\equiv\! \Sigma_{23}-\Sigma_{12},
\delta\Sigma_{33}\!\equiv\! \Sigma_{33}+\Sigma_{11}-2\Sigma_{13}
\ee
We then expand these $ \delta \Sigma $ after recalculation of their integrand by linear combination of the integrands of $ \Sigma_{ij} $. It suffices here to know the dominant order of $ \delta \Sigma $ and of the remaining $ \Sigma_{ij} $, except for $ \delta \Sigma_{33} $ where the subleading order is required: 
\bea
\label{eq:017}
&\hspace{-5mm}&\check{\Sigma}_{11}^{[-1]}=\frac{\check{\Delta}}{8\ii\pi} \asin \frac{1}{\sqrt{\zeta}}\,,\ \check{\Sigma}_{22}^{[1]}=\frac{\zeta\asin\frac{1}{\sqrt{\zeta}} + \sqrt{\zeta-1}}{16\ii\pi\check{\Delta}}\,,\nonumber\\
&\hspace{-5mm}& \check{\Sigma}_{12}^{[0]}=\frac{\sqrt{\eee^{2\tau}\!-\!1}}{-(2\pi)^2} \left[\re\Pi(\eee^{\tau},\ii \eee^{\tau})\!-\!\Pi(-\eee^{\tau},\ii \eee^{\tau})\!+\!\frac{K(\ii\eee^{\tau})}{\sh\tau}\right]\,,\nonumber \\
&\hspace{-5mm}& \delta\check{\Sigma}_{13}^{[1]}=\frac{\ii\sqrt{\zeta-1}}{16\pi\check{\Delta}}\,,\ \delta\check{\Sigma}_{23}^{[2]}=\frac{(2/3-\zeta)}{2\check{\Delta}^2}\check{\Sigma}_{12}^{[0]}-\frac{\sqrt{1\!-\!\eee^{-2\tau}}}{24\pi^2}\times \nonumber\\
&\hspace{-5mm}& [E(\ii\eee^\tau)\!-\!\eee^\tau\ch\tau K(\ii\eee^\tau)]\,,\ \delta\check{\Sigma}_{33}^{[2]}=\frac{\sqrt{1\!-\!\eee^{-2\tau}}}{24\pi^2\check{\Delta}}\times [E(\ii\eee^\tau)\nonumber\\
&\hspace{-5mm}& +\coth\tau\, K(\ii\eee^\tau)]\,,\ \delta\check{\Sigma}_{33}^{[3]}=\frac{(\zeta\!-\!2)\sqrt{\zeta\!-\!1}+\zeta^2\asin\frac{1}{\sqrt{\zeta}}}{64\ii\pi\check{\Delta}^3}
\eea
where $ \check {\Sigma} _{ij}^{[n]} $ ($ \delta \check {\Sigma} _{ij}^{[n]} $) is the coefficient of $ \check {q}^n $ in the Taylor expansion of $ \check {\Sigma} _{ij} $ ($ \delta \check {\Sigma} _{ij} $). The first three identities are already in \cite{PRL2019, CRAS2019}. The Czech accent indicates the rescaling of energies by $ \mu $ ($ \check {\Delta} = \Delta / \mu $), wave numbers by $ k_0 $ ($ \check {q} = q / k_0 $) and $ \Sigma_{ij} $ by $ k_0^3 / \mu $, with $k_0=(2m\mu)^{1/2}/\hbar$ here. The expressions of several integrals over $ k $ have been used in terms of complete elliptic integrals $ K $, $ E $ and $ \Pi $ of first, second and third kinds \cite{GR}, in particular those given in reference \cite{Strinati}, after putting $ \sh \tau = 1 / \check {\Delta} $ to abbreviate.\footnote {We also used, for $ x \geq 0 $, $ E (\ii x ) = \sqrt {1 \! + \! x^2} E (x / \sqrt {1 \! + \! x^2}) $ and $ K (\ii x) = K (x / \sqrt { 1 \! + \! X^2}) / \sqrt {1 \! + \! X^2} $ \cite{GR}. For example, $ \int_0^{+ \infty} \! \dd \check {k} \frac {\check {k}^2 \check {\xi} _k} {\check {\varepsilon} _k^3} = K (\ii \eee^\tau) \sqrt {\eee^{2 \tau} \! - \! 1} / 2 $.} We finally get the low $ q $ behavior of response functions: 
\bea
\label{eq:018a}
&\hspace{-5mm}&\check{\chi}_{\rho\rho}\stackrel{\nu\,\mbox{\scriptsize fixed}}{\underset{q\to 0}{=}} 2\check{q}^2\delta\check{\Sigma}_{33}^{[2]} +2\check{q}^3\bigg[\delta\check{\Sigma}_{33}^{[3]} \nonumber \\
&\hspace{-5mm}& +\frac{2\check{\Sigma}^{[0]}_{12}\delta\check{\Sigma}_{13}^{[1]}\delta\check{\Sigma}_{23}^{[2]}-\check{\Sigma}_{22}^{[1]}\delta\check{\Sigma}_{13}^{[1]2}-\check{\Sigma}_{11}^{[-1]}\delta\check{\Sigma}_{23}^{[2]2}}{\check{\Sigma}_{11}^{[-1]}\check{\Sigma}_{22}^{[1]}-\check{\Sigma}_{12}^{[0]2}}\bigg]\!+\!O(\check{q}^4)\\
&\hspace{-5mm}&\chi_{|\Delta|\rho}\stackrel{\nu\,\mbox{\scriptsize fixed}}{\underset{q\to 0}{=}}\check{q} \frac{\check{\Sigma}_{11}^{[-1]}\delta\check{\Sigma}_{23}^{[2]}-\check{\Sigma}_{12}^{[0]}\delta\check{\Sigma}_{13}^{[1]}}{\check{\Sigma}_{11}^{[-1]}\check{\Sigma}_{22}^{[1]}-\check{\Sigma}_{12}^{[0]2}}\!+\!O(\check{q}^2)
\label{eq:018b}
\eea
where $ \chi _{\rho \rho} $ is expressed in units of $ k_0^3 / \mu $ and $ \chi_{| \Delta | \rho} $ is naturally dimensionless. A more explicit dependence on the reduced frequency $ \nu $ is obtained by passing to the limit $ \eta \to 0^+ $ as in equation (\ref {eq:014}): 
\bea
\nonumber
\asin\frac{1}{\sqrt{\zeta}} &\underset{\eta\to 0^+}{\to}& \left\{
\begin{array}{ll}
-\ii\argsh \frac{1}{\sqrt{-\nu}} & \mbox{if }\nu<0 \\
\frac{\pi}{2}-\ii\argch \frac{1}{\sqrt{\nu}} & \mbox{if }0<\nu<1 \\
\asin\frac{1}{\sqrt{\nu}} & \mbox{if }1<\nu
\end{array}
\right.\ \mbox{and} \\
\sqrt{\zeta-1} &\underset{\eta\to 0^+}{\to}& \left\{
\begin{array}{ll}
\ii\sqrt{1-\nu} & \mbox{if }\nu<1 \\
\sqrt{\nu-1} & \mbox{if }\nu>1 
\end{array}
\right.
\eea
It allows to check that the coefficient $ \check {\chi} _{\rho \rho}^{[3]} $ of the contribution of order $ \check {q}^3 $ in (\ref {eq:018a}) and that $ \check{\chi}_{| \Delta | \rho}^{[1]} $ of the contribution of order $ \check {q} $ in (\ref {eq:018b}) have a zero imaginary part for $ \nu <0 $ (this was predictable and happens in the response functions to all orders in $ q $ because the density of states of the pair-breaking continuum $\kk\mapsto\varepsilon_{\qq/2+\kk}+\varepsilon_{\qq/2-\kk}$ is zero at energies below $2\Delta$) and a zero real part  for $ \nu> 1 $ (this for a physical reason that we do not know, and which does not hold to all orders in $q$). These coefficients have a finite and real limit at $ \nu = 0 $, reached slowly (logarithmically, with a deviation varying as $ 1 / \ln | \nu | $), 
\bea
\label{eq:021a}
\lim_{\nu\to 0} \check{\chi}_{\rho\rho}^{[3]}(\nu) &=& -\frac{1}{16\pi\check{\Delta}^3} -32\pi\check{\Delta} [\delta\check{\Sigma}_{23}^{[2]}(\nu=0)]^2 \\
\label{eq:021b}
\lim_{\nu\to 0} \check{\chi}_{|\Delta|\rho}^{[1]}(\nu) &=& 16\pi\check{\Delta}\,\delta\check{\Sigma}_{23}^{[2]}(\nu=0)
\eea
which gives rise to a sharp spike, with a vertical tangent, in the $ \nu $ dependence, as in reference \cite{PRL2019}; they have at $ \nu = 1 $ a singularity $ | \nu-1 |^{1/2} $, on the real part for $ \nu \to 1^- $, on the imaginary part for $ \nu \to 1^+ $, which gives rise to an ordinary kink with a vertical tangent (see figure \ref {fig:gra} to come). 

\begin{figure} [t] 
\centerline {\includegraphics [width = 4.25cm, clip =] {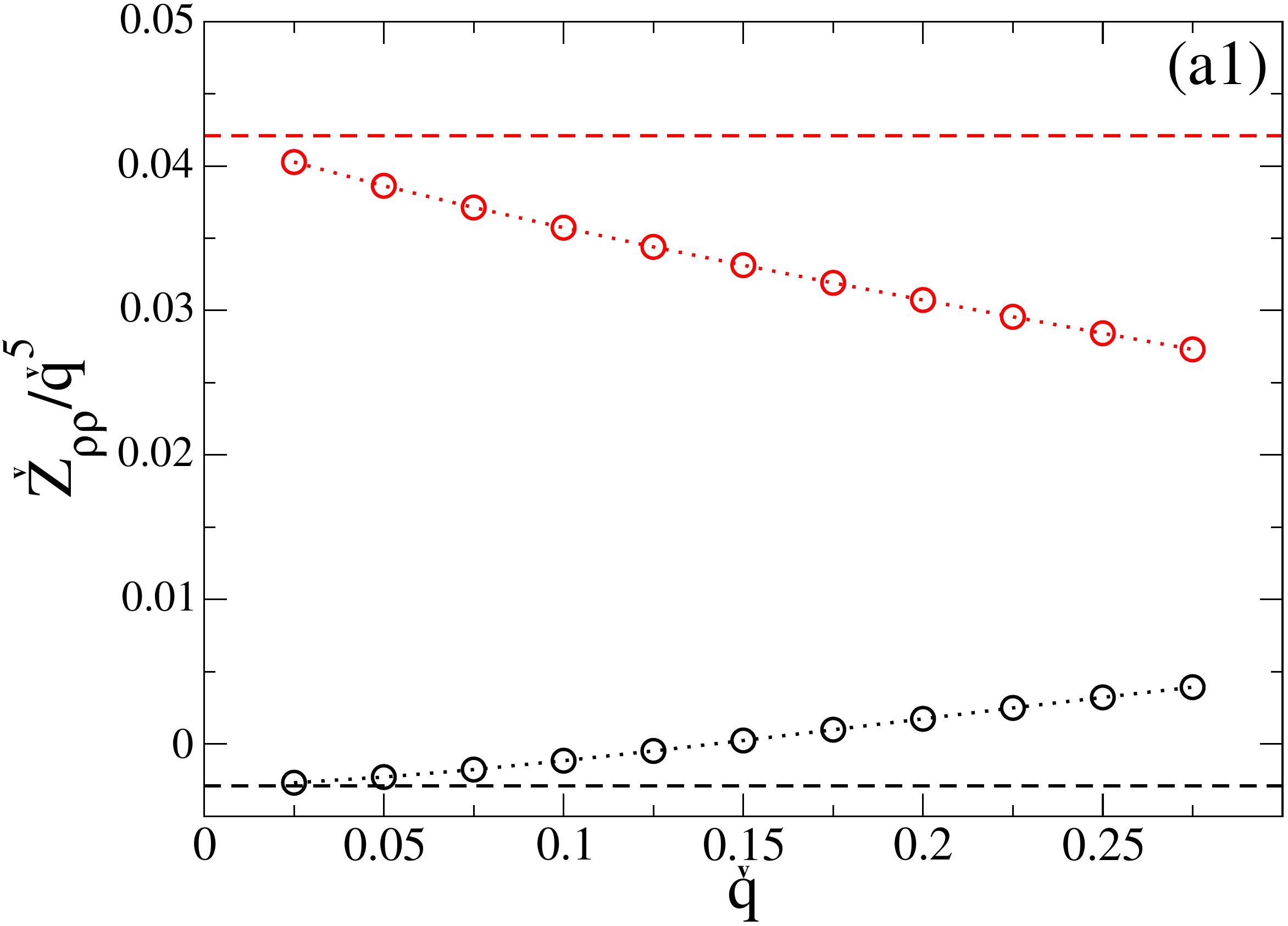} \hspace {3mm} \includegraphics [width = 4.25cm, clip =] {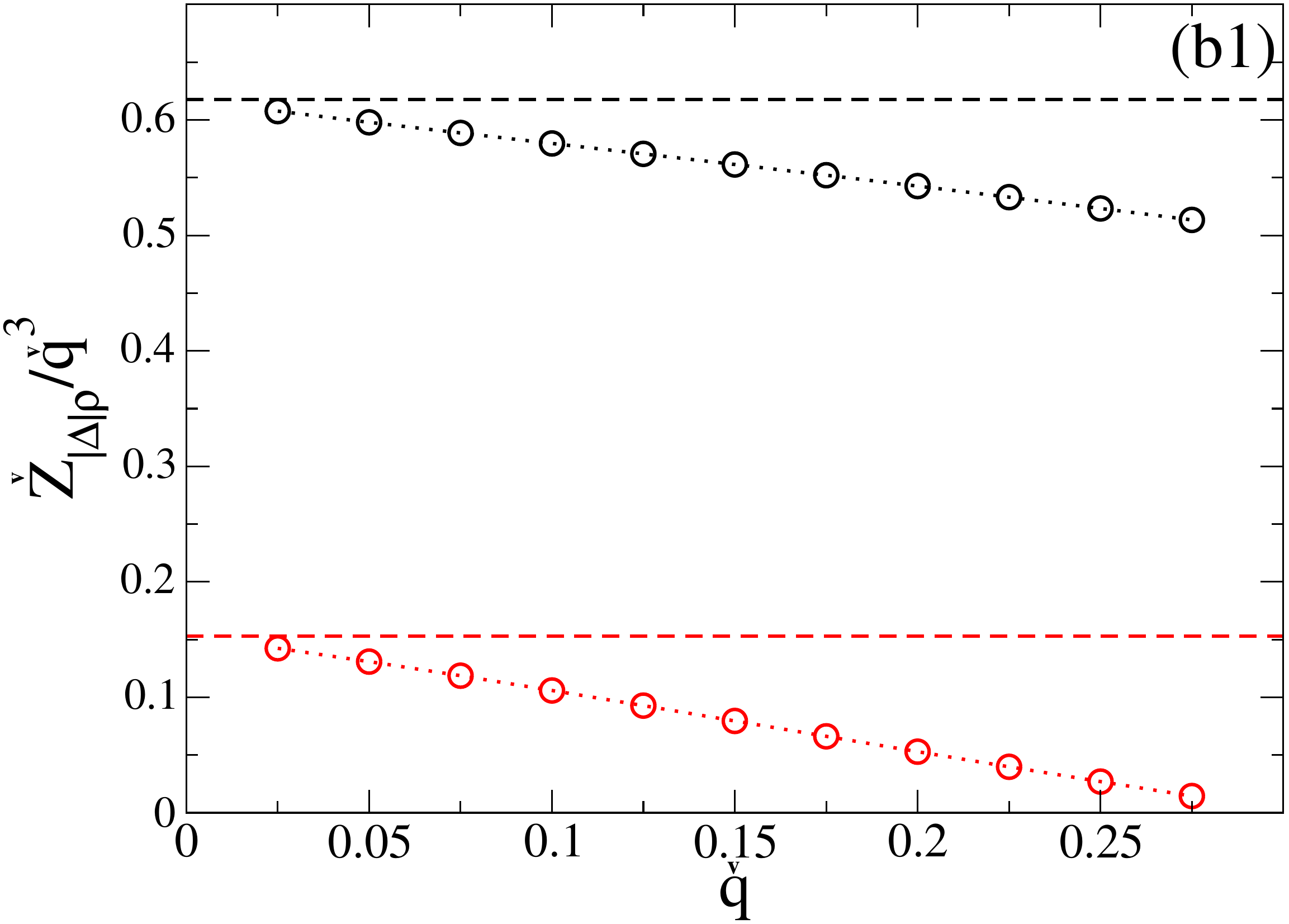}} 
\centerline {\includegraphics [width = 4.25cm, clip =] {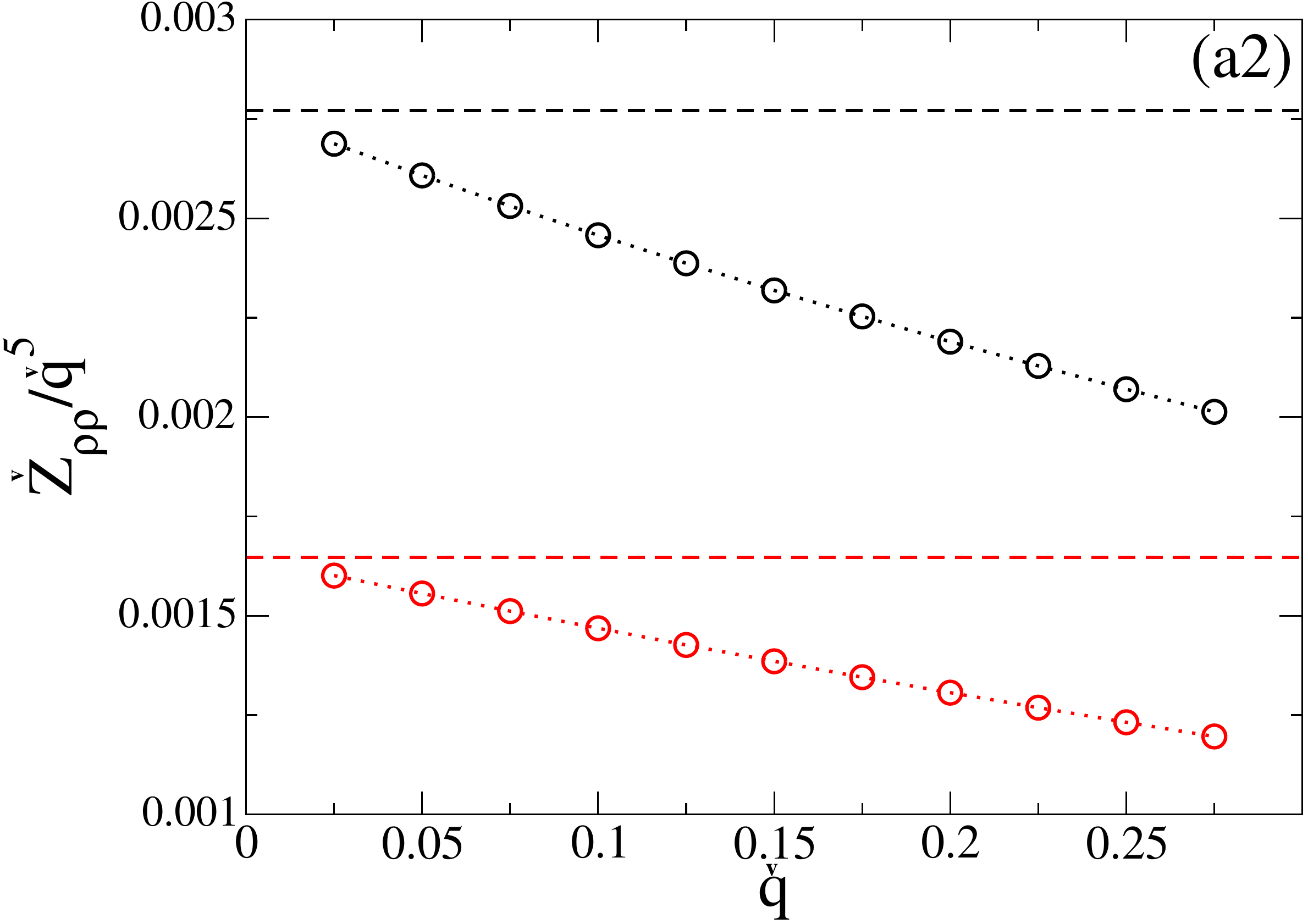} \hspace {3mm} \includegraphics [width = 4.25cm, clip =] {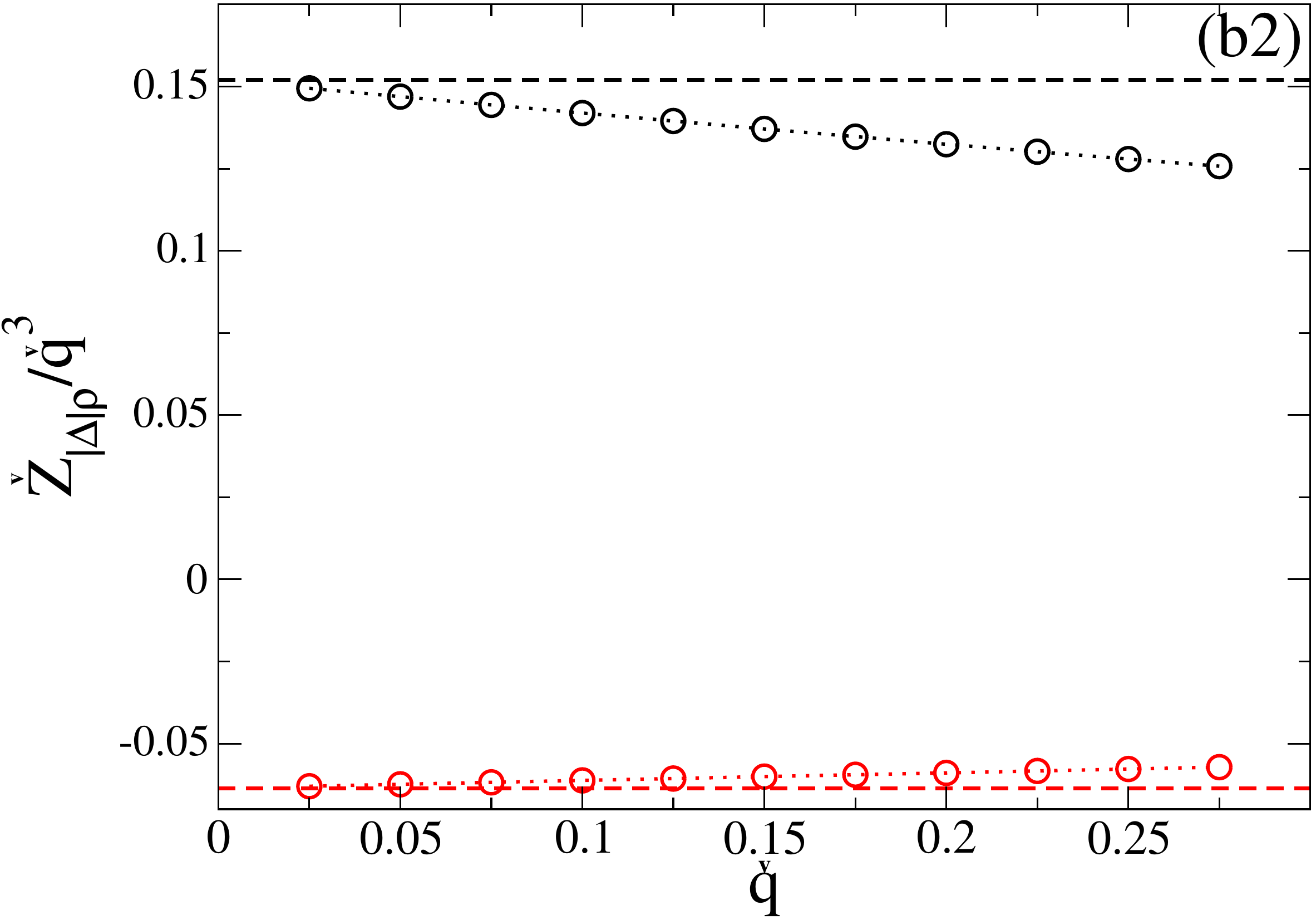}} 
\caption {Complex spectral weight of the continuum mode in the density-density (column a) and modulus-density (column b) response functions, that is the residue $ Z _{\rho \rho} $ or $ Z_{| \Delta | \rho} $ of the analytic continuation of $ \chi _{\rho \rho} (\qq, z / \hbar) $ or of $ \chi_{| \Delta | \rho} (\qq, z / \hbar) $ from $ \im z> 0 $ to $ \im z <0 $ (through the interval between their first two singularities $\varepsilon_1(q)$ and $\varepsilon_2(q)$ on $ \mathbb {R}^+ $) at the pole $ z_\qq $ (complex energy of the mode), as functions of the wave number $ q $, on the $ \mu> 0 $ side of the BEC-BCS crossover (\ref {eq:010}), for $ \check {\Delta} = 1/2 $ (line 1) and $ \check {\Delta} = 2 $ (line 2). The residues were divided by the power of $ q $ ensuring the existence of a finite and nonzero limit at $ q = 0 $. In black: real part; in red: imaginary part. Circles connected by a dotted line: numerical results from general forms (\ref {eq:011}); the analytic continuation is carried out as in \cite{PRL2019, CRAS2019} by the method of the spectral densities of reference \cite{Noz}, see our footnote \ref {note:pro2}. Horizontal dashes: limit at $ q = 0 $, taken from the analytical results (\ref {eq:018a}, \ref {eq:018b}) continuated as in footnote \ref {note:pro}. Czech accent: $ \Delta $ rescaled by $ \mu $ ($ \check {\Delta} = \Delta / \mu $), $ q $ rescaled by $ k_0 = (2m \mu)^{1/2 } / \hbar $ ($\check{q}=q/k_0$), $ Z _{\rho \rho} $ rescaled by $ k_0^3 $ and $ Z_{| \Delta | \rho} $ rescaled by $ \mu $. } 
\label {fig:res} 
\end {figure}

\begin {figure} [t] 
\centerline {\includegraphics [width = 4.25cm, clip =] {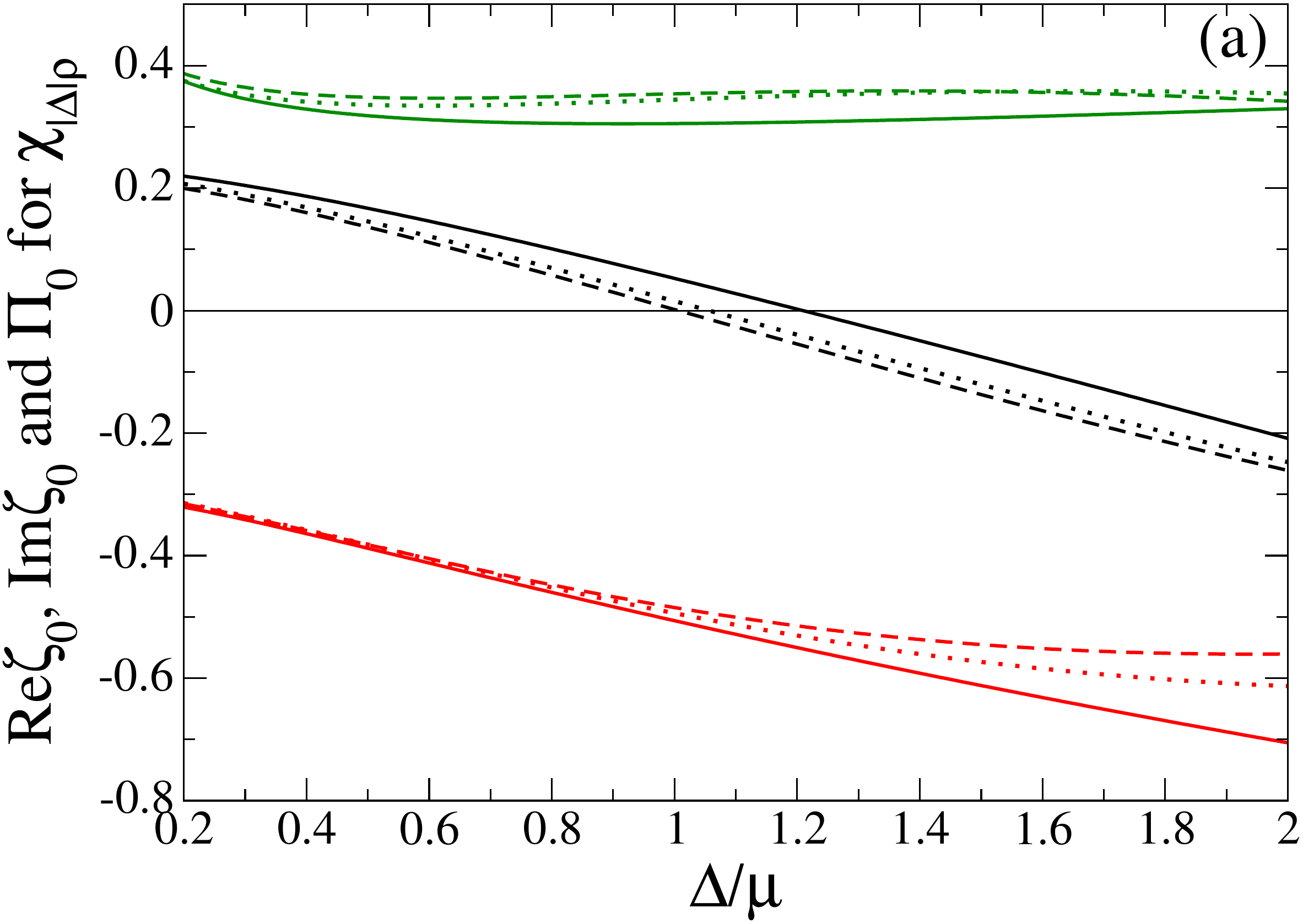} \hspace {3mm} \includegraphics [width = 4.25cm, clip =] {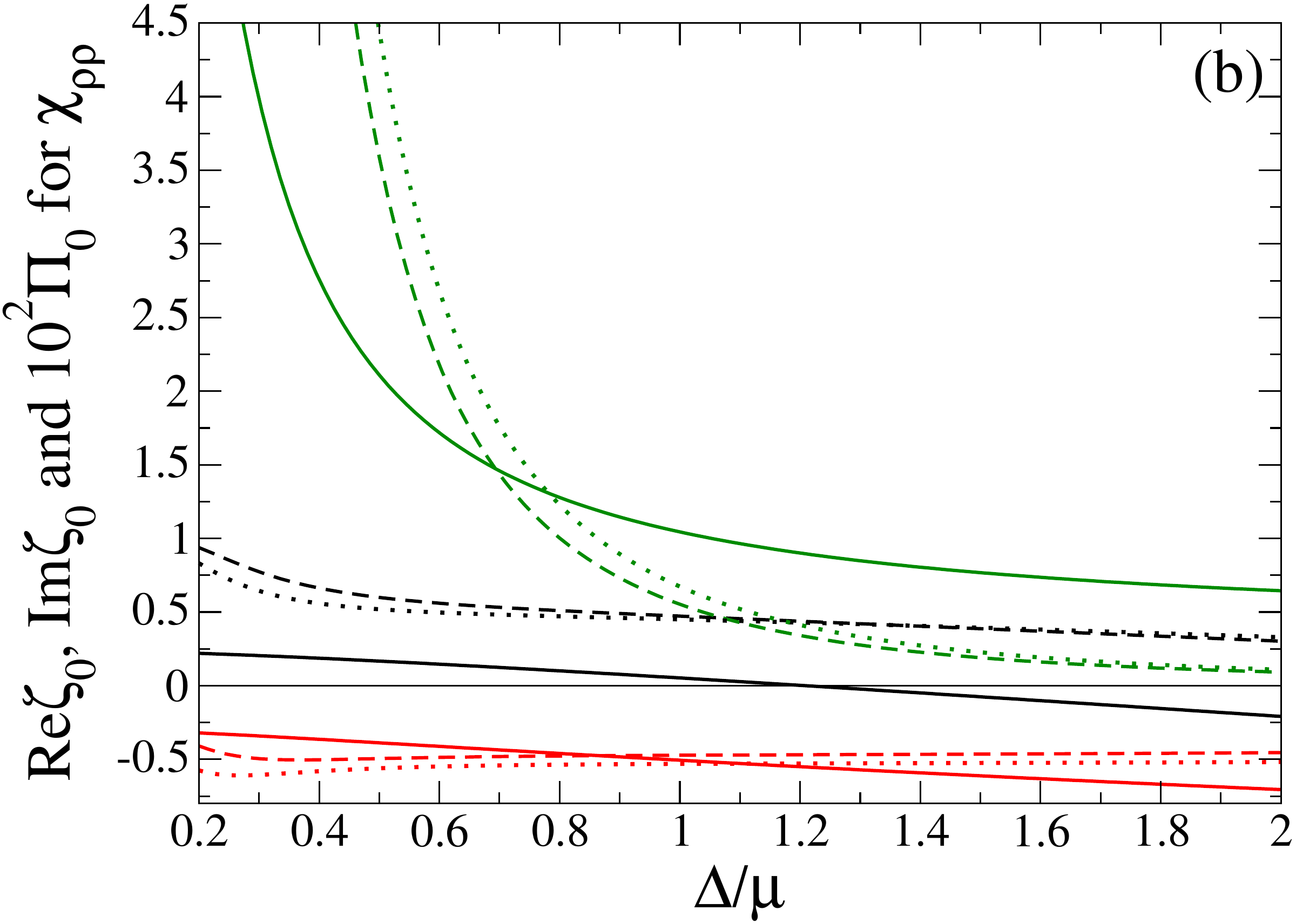} }
\caption {Extraction of the complex energy of the continuum mode and its spectral weight by a fit with four complex parameters (\ref {eq:022}) of the response functions $ \chi (\qq, \omega) $ in the BEC-BCS crossover on the $ \mu> 0 $ side (the value $\check{\Delta}=0.2$ corresponding to $k_{\rm F}a\simeq -1$ in BCS theory). (a) By fitting the coefficient $ \check {\chi}^{[1]} _{| \Delta | \rho} (\nu) $ of $ \check {q} $ in the expansion (\ref {eq:018b}) of the modulus-density response function. (b) Idem for the coefficient $ \check {\chi}^{[3]} _{\rho \rho} (\nu) $ of $ \check {q}^3 $ in (\ref {eq:018a}). In black (red): real (imaginary) part of the coefficient $ \zeta_0 $ in the quadratic start (\ref {eq:012}) of the complex energy. In green: spectral weight $ \Pi_0 $ of the mode in the considered response function (this is the residue modulus of the analytic continuation of $ \check {\chi}^{[n]} $ at $ \zeta_0 $, so that $ \Pi_0 = \lim _{\check {q} \to 0} \check {\Delta} | \check {Z} _{| \Delta | \rho} | / \check {q}^3 $ in (a) and $ \Pi_0 = \lim _{\check {q} \to 0} \check {\Delta} | \check {Z} _{\rho \rho} | / \check {q}^5 $ in (b), where the $ \check {Z} $ are those of figure \ref {fig:res} and account was taken of rescaling (\ref{eq:013})). Full line: exact values. Dashed line: values taken from the fit on the reduced frequency interval $ \nu \in [1 / 10,9 / 10] $. Dotted line: same for $ \nu \in [1/5, 4/5] $. The fit interval was discretized at 60 equally spaced points. Note the factor $ 100 $ on $ \Pi_0 $ in (b).} 
\label {fig:fit} 
\end {figure}

Let's physically analyze results (\ref {eq:018a}, \ref {eq:018b}). First, the dominant term (of order $ q^2 $) in the density-density response function is of little interest for our study: it is insensitive to the continuum mode since the functions $ \Sigma_{ij} (z, \qq) $, even after continuation to the lower half complex plane, have no poles. Fortunately, it constitutes a background independent of the reduced frequency $ \nu $, as can be verified on equation (\ref {eq:017}); it is therefore possible to get rid of it experimentally by considering the difference 
\be
\label{eq:019}
\check{\chi}_{\rho\rho}(\check{q},\nu)-\check{\chi}_{\rho\rho}(\check{q},\nu_0)
\ee
where $ \nu $ is the running variable and $ \nu_0 $, the reduced reference frequency, is fixed. We also note that this background of order $ q^2 $ is real, so that it does not contribute to the imaginary part of $ \chi_{\rho \rho} $, which is often what we measure really in the experiment \cite{Bragg5}. On the other hand, the subdominant term (of order $ q^3 $) in $ \chi _{\rho \rho} $ is sensitive to the continuum mode: as it contains functions $ \Sigma_{ij} $ in the  denominator, its analytic continuation to the lower half complex plane through the interval $ \nu \in [0,1] $ has a pole at $ \zeta = \zeta_0 $, where the complex number $ \zeta_0 $ is that of equation (\ref {eq:012}), with a nonzero residue, see figure \ref {fig:res}a. The same conclusion applies for the dominant term (of order $ q $) in the modulus-density response function, see figure \ref {fig:res}b.\footnote {\label {note:pro} The analytic continuation of $ \zeta \mapsto \check {\chi} _{\rho \rho}^{[3]} $ and $ \zeta \mapsto \check{\chi}_{| \Delta | \rho}^{[1 ]} $ from the upper half-plane to the lower half-plane is done by passing through the interval $ [0,1] $ (connecting their singularities in $ \nu = 0 $ and $ \nu = 1 $ on the real axis) by the substitutions $ \asin \frac {1} {\sqrt {\zeta}} \to \pi-\asin \frac {1} {\sqrt {\zeta}} $ and $ \sqrt {\zeta -1} \to - \sqrt {\zeta-1} $ as in reference \cite{PRL2019}. The analytic continuation of the denominator in the right-hand side of equations (\ref {eq:018a}) and (\ref {eq:018b}) gives precisely the function of \cite{PRL2019} whose $ \zeta_0 $ is a root. There is no other continuation interval to consider because the denominator of $ \check {\chi} _{\rho \rho}^{[3]} $ and $ \check{\chi}_{| \Delta | \rho }^{[1]} $ (continuated to $ \mathbb {C} \setminus \mathbb {R} $ by the identities $ \Sigma_{ij} (z) = [\Sigma_{ij} (z^*)]^* $) has no branch cut for $ \nu \in] - \infty, 0] $ (due to a vanishing density of states of the pair-breaking continuum) nor for $ \nu \in [1, + \infty [$ (due to a compensation of the discontinuities of $ \check {\Sigma} _{11}^{[- 1]} $ and $ \check {\Sigma} _{22}^{[1]} $ on this half-line, which are simply sign changes).}\ \footnote {\label {note:pro2} The analytic continuation of the functions $ \Sigma_{ij} (z) $ from $ \im z> 0 $ to $ \im z <0 $ is given by $ \Sigma_{ij}\!\!\downarrow\!\!(z) = \Sigma_{ij} (z) - \frac {2 \ii \pi} { (2 \pi)^3} \rho_{ij} (z) $ in terms of the spectral densities defined on $ \mathbb {R}^+ $ by $ \im \Sigma_{ij} (\varepsilon + \ii 0^+ ) = - \frac {\pi} {(2 \pi)^3} \rho_{ij} (\varepsilon) $ \cite{Noz}, which it suffices to know on the continuation interval between the first two branching points $\varepsilon_1(q)=2\Delta$ and $ \epsilon_2 (q) $ \cite{PRL2019}. Then $ \rho_{13} (\varepsilon) = (\frac {2m} {\hbar^2})^2 \frac {\pi \varepsilon} {2 q} K (\ii \sh \Omega) $, $ \rho_{23} (\varepsilon) = 0 $, $ \rho_{33} (\varepsilon) = (\frac {2m} {\hbar^2})^2 \frac {\pi \Delta} {q } E (\ii \sh \Omega) $, with $ \Omega = \argch \frac {\varepsilon} {2 \Delta} $. The other $ \rho_{ij} (\varepsilon) $ are in references \cite{PRL2019, CRAS2019}. }
Unsurprisingly, the residue $Z$ of the continuum mode, plotted in these figures as a function of $ q $, is complex, since both the response functions and the energy $ z_ \qq $ are complex. In practice, the phase of $ Z $ does not matter (the pole is unique and its contribution cannot interfere with that of another pole) and it is its modulus which characterizes the spectral weight of the mode; we therefore plot $ | Z | $ (or more precisely the coefficient of its dominant order in $ q $) as a function of the interaction strength in figure \ref{fig:fit}.

However, the physical measurements take place on the real axis of angular frequencies. So we have plotted $ \check {\chi}^{[3]} _{\rho \rho} $ and $ \check{\chi}_{| \Delta | \rho}^{[1]} $ as functions of the reduced frequency $ \nu $ on figure \ref {fig:gra}, for two values of the interaction strength. The expected narrow structures should be on the analytic continuation interval $ \nu \in [0,1] $, almost above the pole $ \zeta_0 $ so close to the vertical green solid line. For $ \check {\Delta} = 1/2 $ (first line of the figure), we are in the favorable case $ \re \zeta_0 \in [0,1] $; now, $ \check {\chi}^{[3]} _{\rho \rho} $ exhibits, on the interval $ [0,1] $, a shoulder-shaped structure with a maximum and a minimum, as well on its real part as on its imaginary part; even better, $ \check {\chi}^{[1]} _{| \Delta | \rho} $ has a fairly pronounced bump on its real part over the same interval, even if it is quite far from the green line, and a fairly visible dip on its imaginary part, close to the line. For $ \check {\Delta} = 2 $ (second line of the figure), we are in the disadvantageous case $ \re \zeta_0 <0 $; the response functions should therefore keep track of the continuum mode over the interval $ \nu \in [0,1] $ only through the associated complex resonance wing, and no longer as extrema; unfortunately, the structures observed remain essentially the same as for $ \check {\Delta} = 1/2 $, which casts a dreadful doubt on their link with the continuum mode.\footnote {When $ \re \zeta_0 <0 $ in general, one should not expect to see any bump or dip associated with the continuum mode in the response functions on the interval $ \nu \in] - \infty, 0 [$. Indeed, this physical interval is separated from the pole by the end of the $ [0,1] $ branch cut that had to be folded back to $] - \infty, 0] $ to carry out the analytic continuation, the other end being folded back on $ [1, + \infty [$.} 

\begin {figure} [t] 
\centerline {\includegraphics [width = 4.25cm, clip =] {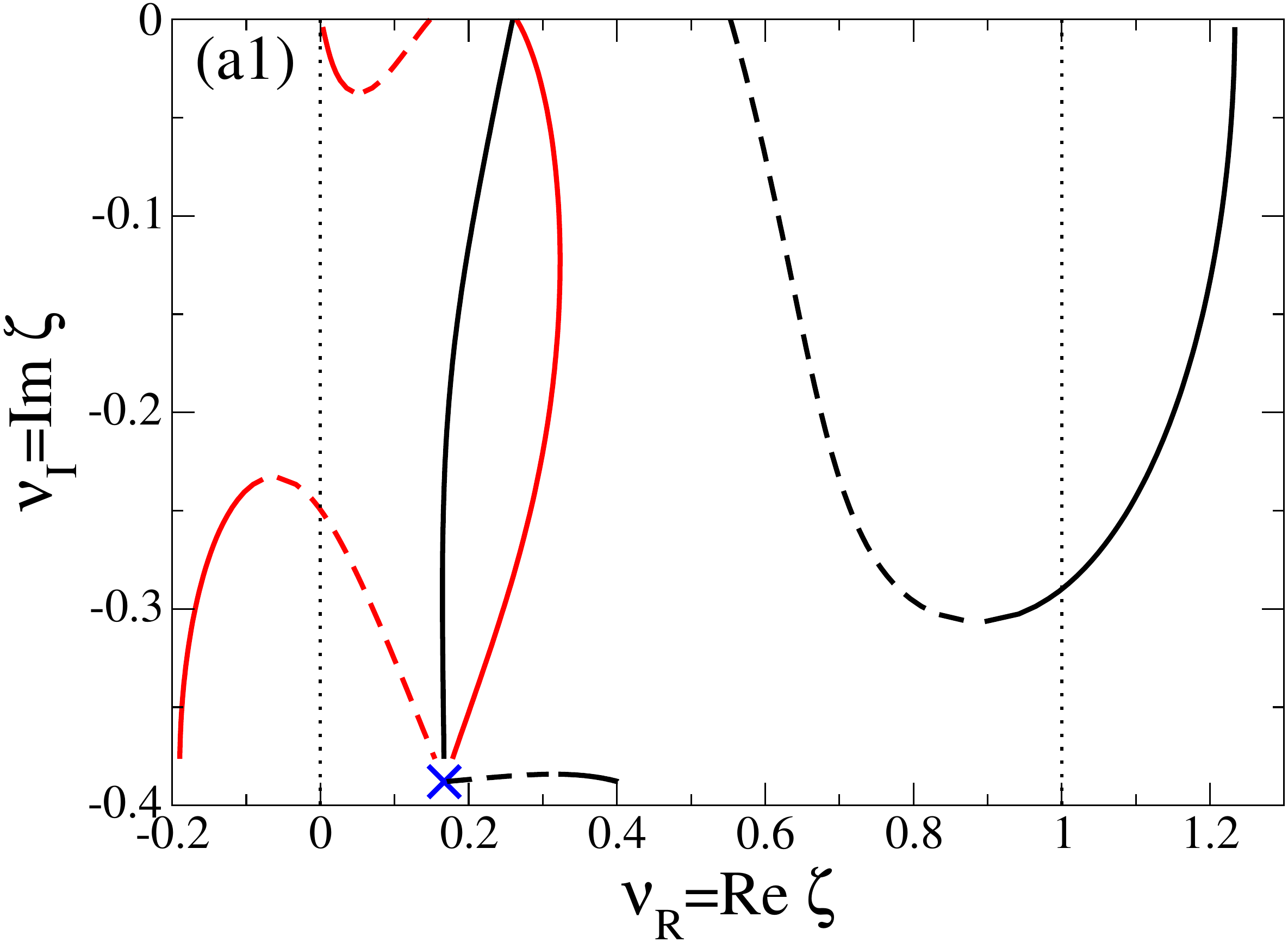} \hspace {3mm} \includegraphics [width = 4.25cm, clip =] {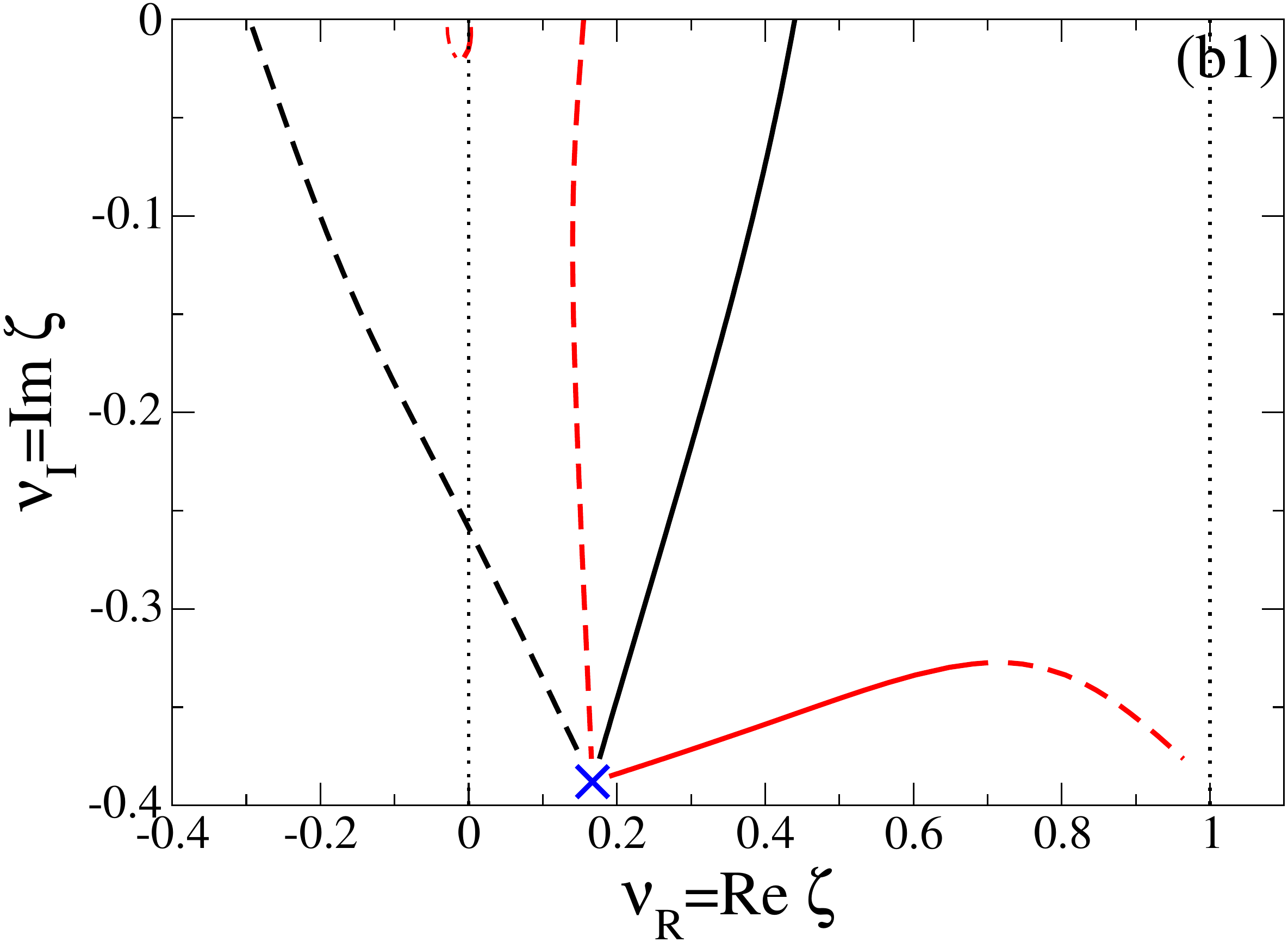}}
\centerline {\includegraphics [width = 4.25cm, clip =] {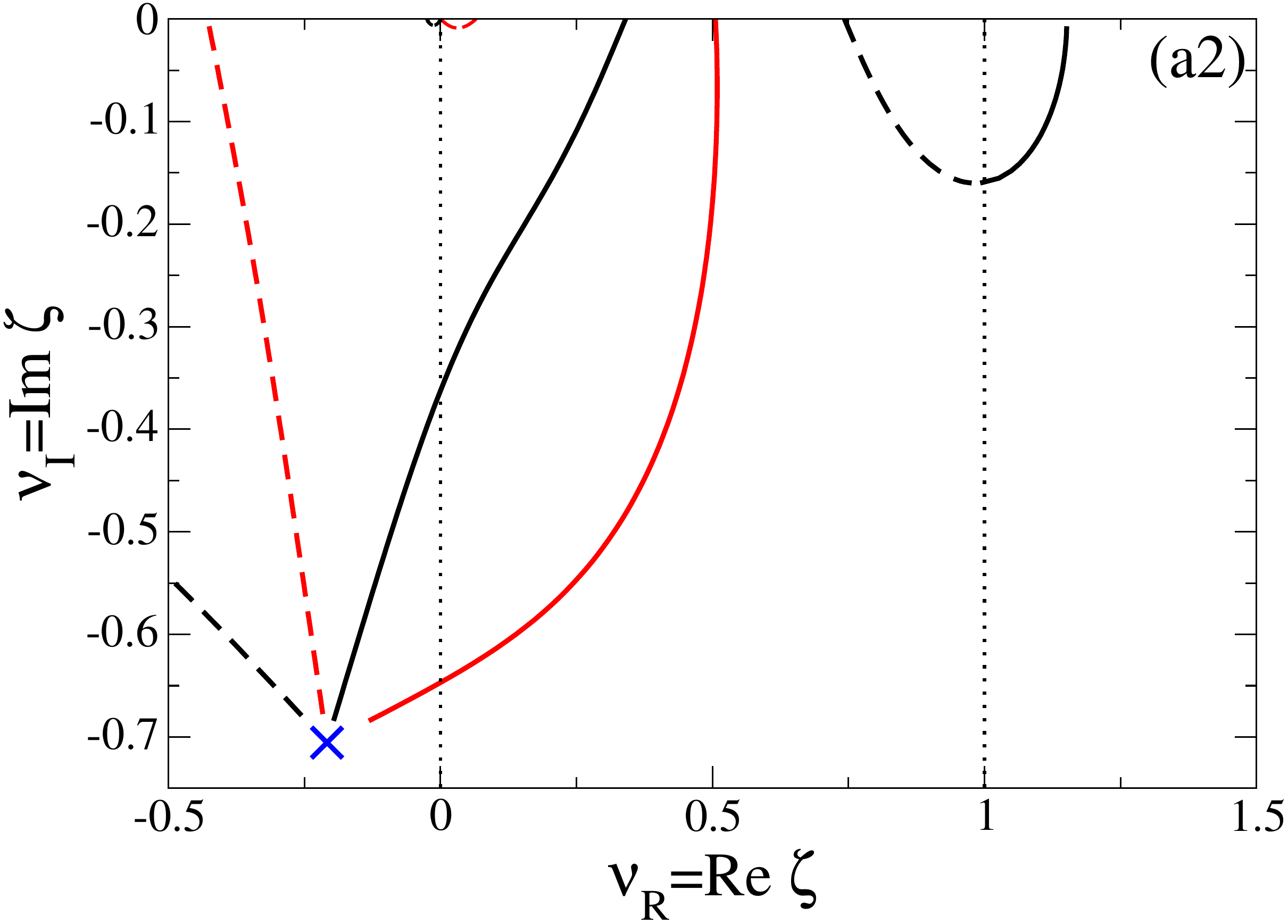} \hspace {3mm} \includegraphics [width = 4.25cm, clip =] {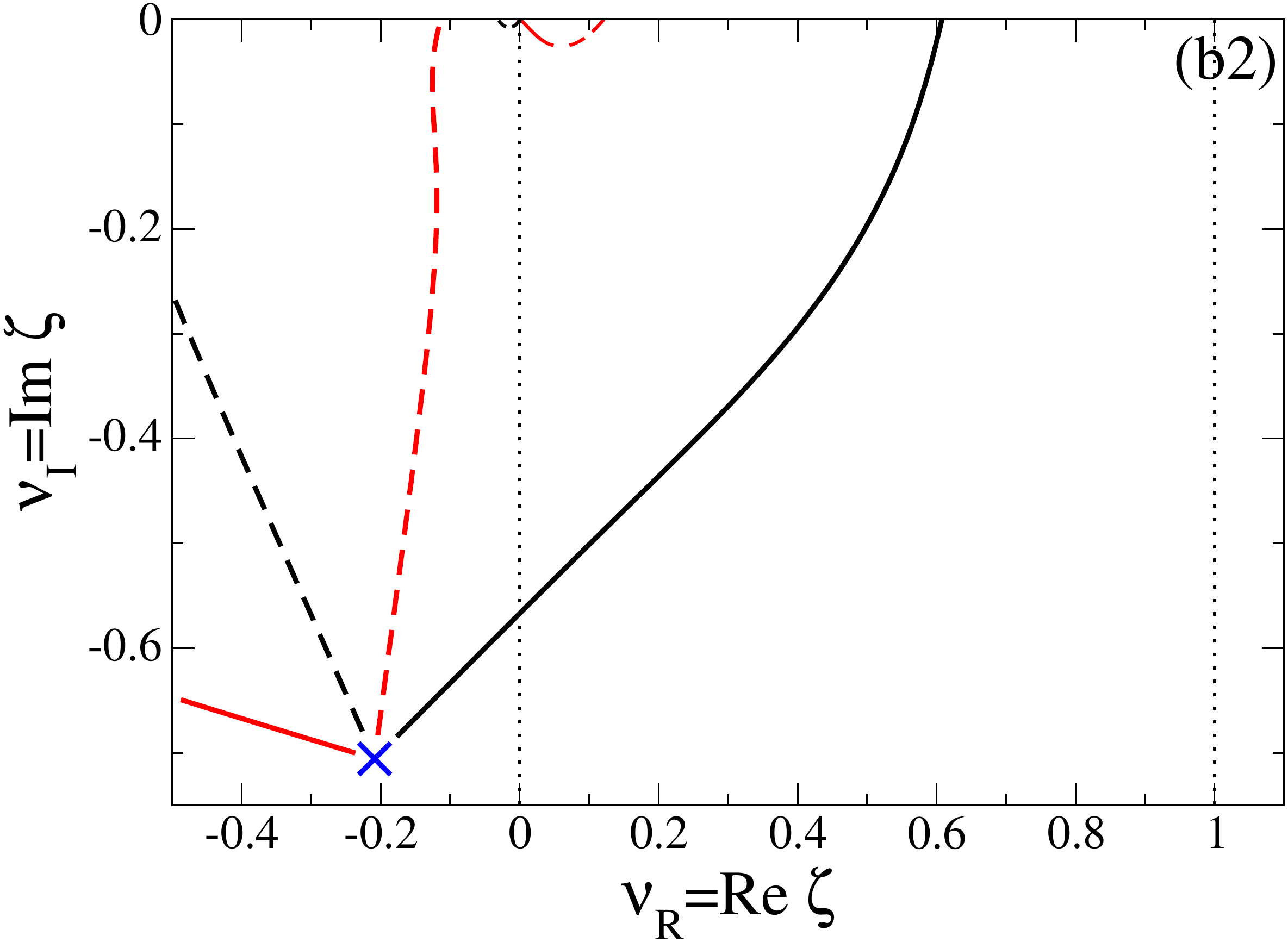}} 
\caption {Locus of the extrema of the functions of the real part $ \nu_R \mapsto \re \chi^{[n]} \! \! \! \downarrow \! \! (\zeta = \nu_R + \ii \nu_I) $ (in black) and $ \nu_R \mapsto \im \chi^{[n]} \! \! \! \downarrow \! \! (\zeta = \nu_R + \ii \nu_I) $ (in red) when the imaginary part $ \nu_I $ of $ \zeta $ varies. Here $ \chi^{[n]} \! \! \! \downarrow \! \! (\zeta) $ is the coefficient of $ q^n $ in the low $ q $ expansion of the susceptibility $ \chi(\qq, \omega) $ at fixed $ \nu $ in (\ref{eq:013}), see equations (\ref{eq:018a}, \ref {eq:018b}), analytically continuated from $ \im \zeta> 0 $ to $ \im \zeta <0 $ through $ \nu \in [0,1] $ as indicated by the arrow $ \downarrow $ in the notation $ \chi \! \! \downarrow$. Column (a): $ \chi = \chi _{\rho \rho} $ and $ n = 3 $; column (b): $ \chi = \chi_{|\Delta| \rho} $ and $ n = 1 $. The fermion gas is in the BEC-BCS crossover on the $ \mu> 0 $ side: $ \Delta / \mu = 1/2 $ (line 1) and $ \Delta / \mu = 2 $ (line 2). Full line: the extremum is a maximum. Dashed line: the extremum is a minimum. Blue cross: reduced complex energy $ \zeta_0 $ of the continuum mode. Vertical dots: positions $ \nu = 0 $ and $ \nu = 1 $ of the singularities of $ \chi^{[n]} (\nu) $ on the real axis.} 
\label {fig:lie} 
\end {figure}

To see what really happens, we perform an analytic continuation of the coefficients $ \check {\chi}^{[3]} _{\rho \rho} (\zeta) $ and $ \check{\chi}_{| \Delta | \rho}^{[1]} (\zeta ) $ to the lower complex plane $ \im \zeta <0 $ as in footnote \ref {note:pro}, then we find the position of the extrema of the real part or the imaginary part of these coefficients, that is, their abscissa $ \nu_R $, on the horizontal line $ \zeta = \nu_R + \ii \nu_I $ of fixed ordinate $ \nu_I $, and we finally plot the locus of these extrema when $ \nu_I $ varies in the interval $] \im \zeta_0,0] $, see figure \ref {fig:lie}. This locus is the union of continuous lines (its connected components); some, but not all,\footnote {A line of maxima and a line of minima  of $ \re \chi $ ($ \im \chi $) can meet to a terminal point where $ \re \partial _{\nu_R}^2 \chi \! = \! 0 $ ($ \im \partial _{\nu_R}^2 \chi \! = \! 0 $).} converge to the pole $ \zeta_0 $.\footnote {At $ \zeta_0 $, we generally expect to see a line of minima and a line of maxima of the real part and of the imaginary part converge, i.e.\, four lines in total. Indeed a meromorphic function $ f (\zeta) $ in the neighborhood of its pole $ \zeta_0 $ is equivalent to $ Z / (\zeta- \zeta_0) $, where $ Z $ is the residue. From real and imaginary part decompositions $ Z = a + \ii b $, $ \zeta- \zeta_0 = x + \ii y $, and the change of scale $ x = y X $, where $ y> 0 $ is the distance from the horizontal line $ \zeta = \nu_R + \ii \nu_I $ to the pole, we get $ f (\zeta) \sim y^{- 1} (\frac {a X + b} {X^2 + 1 } + \ii \frac {bX-a} {X^2 + 1}) $. Now, for all $ u \in \mathbb {R} $, the function $ X \mapsto (X + u) / (X^2 + 1) $ has on $ \mathbb {R} $ a minimum at $ -u - \sqrt {1 + u^2} $ and a maximum at $ -u + \sqrt {1 + u^2} $. So, if $ a \neq 0 $ ($ b \neq 0 $), we see converge to $ \zeta_0 $ two lines of extrema of the real (imaginary) part. For $ \check {\Delta} = 1/2 $, the line of minima of $ \re \check {\chi}^{[3]} _{\rho \rho} $ comes almost horizontally (from the right, with a slope $ \simeq -a / 2b $, see figure \ref{fig:lie}a1) because the residue is almost purely imaginary, $ \check {Z} _{\rho \rho} \sim (-0.003 + 0.04 \ii) \check {q}^5 $ as we see in figure \ref {fig:res}a1.} Any extremum of $ \check {\chi}^{[3]} _{\rho \rho} $ or $ \check{\chi}_{| \Delta | \rho}^{[1]} $ on the real interval $ \nu \in] 0,1 [$ connected continuously to the pole by one of these lines is undoubtedly a physical mark of the continuum mode, observable in the associated response function; the other extrema on the real axis are not. Hence the verdict on figure \ref {fig:gra}: for $ \check {\Delta} = 1/2 $, only the maximum of $ \re \check {\chi}^{[3]} _{\rho \rho} $, the maximum of $ \im \check {\chi}^{[3]} _{\rho \rho} $, the maximum of $ \re \check{\chi}_{| \Delta | \rho}^{[1]} $ and the minimum of $ \im \check{\chi}_{| \Delta | \rho}^{[1]} $ on $ \nu \in] 0,1 [$ are physical marks of the continuum mode; for $ \check {\Delta} = 2 $, this is the case only for the maximum of $ \re \check {\chi}^{[3]} _{\rho \rho} $, for the maximum of $ \im \check {\chi}^{[3]} _{\rho \rho} $ and the maximum of $ \re \check{\chi}_{| \Delta | \rho}^{[1]} $ on $ \nu \in] 0,1 [$. 

Ultimately, it remains to be seen how well we can extract the position and the spectral weight of the continuum mode from measures of response functions on the reduced frequency interval $ \nu \in [0, 1] $. To this end, we propose a very simple fit of the susceptibilities $ \chi (\qq, \omega) $ by the sum of the contribution of the collective mode pole and a slowly variable affine background describing the broad response of the continuum: 
\be
\label{eq:022}
\check{\chi}_{|\Delta|\rho}^{[1]}(\nu)|_{\rm fit} = \frac{A}{\nu-B} +C +D\nu \quad (A,B,C,D\in\mathbb{C})
\ee
taking the example of the modulus-density response limited to its dominant order in $ q $. The fit function is balanced in its search for precision, since it describes the background with the same number of complex adjustable parameters ($ C $ and $ D $) as the resonance ($ A $ and $ B $), that is one parameter more than in reference \cite{PRL2019}. The fit is performed on a sub-interval $ [\nu_1, \nu_2] $ of $ [0,1] $ in order to avoid the singularities at the boundaries. The result is very encouraging for the modulus-density response, see figure \ref {fig:fit}a: we obtain a good approximation of the complex energy and the spectral weight of the mode, even for $ \check { \Delta}> 1.21 $ where $ \re \zeta_0 <0 $ and where the pole is no longer below the analytic continuation (and measurement) interval $ \nu \in [0,1] $. On the other hand, the result is bad for the density-density response, see figure \ref {fig:fit}b, except perhaps for the width of the resonance. To understand this difference in success according to the observable $ | \Delta | $ or $ \rho $, we have calculated the relative height $ h _{\rm rel} $ of the contribution of the resonance above the background.\footnote{For a given function $ \chi (\nu) $, we define the background by $ F (\nu) = \chi (\nu) -Z_0 / (\nu- \zeta_0) $, $ \zeta_0 $ being the pole of the analytic continuation of $ \chi $ to the lower complex half-plane and $ Z_0 $ the associated residue. Then $ h _{\rm rel} = | Z_0 / (\nu_0- \zeta_0) | / | F (\nu_0) | $ with $ \nu_0 = \max (\nu_1, \re \zeta_0) $.} For $ \check {\Delta} <2 $, we always find that $ h _{\rm rel}> 1 $ for the modulus of the order parameter, but that $ h _{\rm rel} <1 $ for the density. For example, for $ \check {\Delta} = 1/2 $, $ h _{\rm rel}^{| \Delta | \rho} \simeq 1.8 $ while $ h _{\rm rel}^{\rho \rho} \simeq 0.37 $. The problem is that the complex resonance does not emerge from the background enough in the density-density response. This problem becomes prohibitive in the low interaction limit, where $ h _{\rm rel}^{\rho \rho} \to 0 $, while it does not arise for the modulus, since $ h _{\rm rel}^{| \Delta | \rho} \to 2.338 \ldots $ for $ \nu_1 <\re \zeta_0 $, see section \ref {sec: bcs}. 

\section {In the BCS limit of weak interaction} 
\label {sec: bcs} 

The weak attraction regime $ k_F a \to 0^- $, although not very relevant for cold atom experiments, has some theoretical interest: it is indeed there that the used BCS theory is most quantitative and most reliable. A clever way to take the continuous limit of our lattice model corresponds to the chain of inequalities $ 0 <-a \ll b \ll 1 / k_F $: we can continue to replace the integration domain $ \mathcal {D } $ by $ \mathbb {R}^3 $ in the definition of $ \Sigma_{ij} $, but as $ | a | / b \ll 1 $, the interaction between fermions is now in Born's regime of scattering theory, so that $ g_0 $ can be approximated by $ g $ in matrix (\ref {eq:007}), 
\be
\label{eq:023}
g_0 \to g
\ee
and the mean-field Hartree shift does not vanish in the BCS spectrum, with $ \xi_\kk = E_\kk- \mu + \rho g / 2 $. At order one in $ k_F a $, the equation of state of the gas at zero temperature contains precisely this Hartree term, $ \mu = \varepsilon_F + \rho g / 2 $; the corresponding BCS spectrum is simply $ \varepsilon_\kk = [(E_\kk- \varepsilon_F)^2 + \Delta^2]^{1/2} $, in agreement with reference \cite{AndrianovPopov}, and reaches its minimum at the wave number $ k_0 = k_F $. 

Expressions (\ref {eq:011}) of the response functions obtained for $ g_0 = 0 $ are no longer sufficient. We recalculate them starting from general expressions (\ref {eq:009a}, \ref {eq:009b}) and by doing substitution (\ref {eq:023}). Proceeding as in footnote \ref {note:Cramer}, we get\footnote {In Cramer's $ 3 \times $ 3 determinant in the numerator of $ \chi_{|\Delta|\rho} $, we add to the third column the second column multiplied by $ g $ to reduce it to a $ 2 \times 2 $ determinant.} 
\be
\label{eq:025}
\chi_{\rho\rho}=\frac{2\left|\begin{array}{lll} \Sigma_{11} & \Sigma_{12} & \Sigma_{13} \\ \Sigma_{12} & \Sigma_{22} & \Sigma_{23} \\ \Sigma_{13} & \Sigma_{23} & \Sigma_{33} \end{array}\right|}{\det M}\,,\quad
\chi_{|\Delta|\rho} = \frac{\left|\begin{array}{ll} \Sigma_{11} & \Sigma_{13}\\ \Sigma_{12} & \Sigma_{23} \end{array}\right|}{\det M}
\ee
implying that the $ \chi $ are taken at $ (\qq, \omega) $ and the $ \Sigma_{ij} $ are taken at $ (z = \hbar \omega + \ii 0^+, \qq) $. Now, the determinant $ \det M $ is a linear function of the third column vector of $ M $, so that 
\be
\label{eq:026}
\det M =  \left|\begin{array}{ll} \Sigma_{11} & \Sigma_{12}\\ \Sigma_{12} & \Sigma_{22} \end{array}\right|
- g  \left|\begin{array}{lll} \Sigma_{11} & \Sigma_{12} & \Sigma_{13}\\ \Sigma_{12} & \Sigma_{22} & \Sigma_{23} \\ \Sigma_{13} & \Sigma_{23} & \Sigma_{33} \end{array}\right|
\ee
By dividing (\ref {eq:025}) up and down by the first term in the right-hand side of (\ref {eq:026}), we recover the susceptibilities (\ref {eq:011}) obtained for $ g_0 = 0 $, which we note $ \chi^{g_0 = 0} $, and which therefore allow us to write very simply the sought susceptibilities at nonzero $ g_0 = g $: 
\be
\label{eq:027}
\chi_{\rho\rho}= \frac{\chi_{\rho\rho}^{g_0=0}}{1-\frac{g}{2}\chi_{\rho\rho}^{g_0=0}}\,, \quad \chi_{|\Delta|\rho}=\frac{\chi_{|\Delta|\rho}^{g_0=0}}{1-\frac{g}{2}\chi_{\rho\rho}^{g_0=0}}
\ee
The forms (\ref {eq:027}) are typical of RPA theory \cite{Anderson}, to which our linearized time-dependent BCS theory is equivalent up to incoming quantum fluctuations \cite{HadrienThese}. Such forms (but not the explicit expressions we give) already appear in references \cite{crrth2, crrth3}. 

It is therefore easy to resume the study of the response functions in the vicinity of the continuum mode at low-wavenumber, by making $ q $ tend to zero at fixed reduced frequency $ \nu $  as in (\ref {eq:013}). Since $ \chi _{\rho \rho}^{g_0 = 0} (\qq, \omega) $ then changes to second order in $ q $, the denominators in (\ref {eq:027}) can be approximated by $ 1 $ and results (\ref {eq:018a}, \ref {eq:018b}) are transposed directly. Remarkably, the whole discussion at low $ q $ in section \ref {sec: CBE-BCS} is actually independent of the precise value of $ g_0 $ and also applies to the case $ g_0 = g $, excepted of course for the value of the function $ \xi_\kk $ and the BCS spectrum $ \varepsilon_\kk $, as well as the position $ k_0 $ of its minimum. 

In this limit $ k_F a \to 0^- $, the equilibrium order parameter tends exponentially to zero, $ \Delta / \varepsilon_F \sim 8 \eee^{- 2} $ $ \exp (- \pi / 2 k_F | a |) $ according to BCS theory \cite{Randeriavert}. This greatly simplifies our results. Let us give the coefficient of $ \check {q}^3 $ and $ \check {q} $ in the Taylor expansions (\ref {eq:018a}, \ref {eq:018b}) of the response functions to the dominant order in $ \Delta $:\footnote {We use section 4.6.3 of reference \cite{CRAS2019} for $ \Sigma_{12} $ and the known expansion of elliptic integrals for the rest. Expansions (\ref {eq:018a}, \ref {eq:018b}) assume that $ q \xi \ll $ 1 so $ \check {q} \ll \check {\Delta} $, hence the order of the limits $ q \to 0 $ then $ \Delta \to 0 $.} 
\bea
\label{eq:028a}
\check{\chi}_{\rho\rho}^{[3]}(\nu)\underset{\check{\Delta}\to 0}{\sim} \frac{(\zeta-2)\sqrt{\zeta-1}+\zeta^2\asin \frac{1}{\sqrt{\zeta}} -\frac{2(\zeta-1)}{\asin \frac{1}{\sqrt{\zeta}}}}{32\ii\pi \check{\Delta}^3}   \\
\label{eq:028b}
\check{\chi}_{|\Delta|\rho}^{[1]}(\nu)\underset{\check{\Delta}\to 0}{\sim} \frac{2}{\ii\pi}\left[\frac{1+\zeta \ln \frac{\check{\Delta}}{8\eee}}{\zeta\asin \frac{1}{\sqrt{\zeta}} +\sqrt{\zeta-1}} -\frac{\frac{1}{2} \ln \frac{\check{\Delta}}{8\eee}}{\asin \frac{1}{\sqrt{\zeta}}}\right] 
\eea
where the energies are this time in units of $ \varepsilon_F $ ($ \check {\Delta} = \Delta / \varepsilon_F $) and the wave numbers in units of $ k_0 = k_F $ ($ \check {q} = q / k_F $). At this order, unlike the modulus response, the density response no longer has a pole in its analytic continuation, so it bears no trace of the continuum mode;\footnote {We must expand $ \check {\chi} _{\rho \rho}^{[3]} (\zeta) $ to the order $ \check {\Delta}^{- 1} $ to find a pole in its analytic continuation, of residue $ \zeta_0^{(0 )} (1+ \zeta_0^{(0)} \ln \frac {\check {\Delta}} {8 \eee})^2 / (2 \ii \pi^3 \check {\Delta} \sqrt {\zeta_0^{(0)} - 1}) $ with $ \zeta_0^{(0)} \simeq 0.2369-0.2956 \ii $.} moreover, on the open interval $ \nu \in] 0,1 [$, its real (imaginary) part is a purely increasing (decreasing) function of $ \nu $, with no extremum, in contrast with figure \ref {fig:gra}a1. 

\section {Conclusion} 
At zero temperature, in the time-dependent BCS approximation, we calculated the linear response $\chi(\qq,\omega)$ of a superfluid gas of spin $ 1 / 2$ fermions to a Bragg excitation of wave vector $\qq$ and angular frequency $ \omega $, which is a well known excitation technique in cold atom experiments. For a chemical potential $ \mu> 0 $, we investigated this response analytically within the low wavenumber limit $ q \to 0 $, the deviation of $ \hbar \omega $ from the edge $ 2 \Delta $ of the pair-breaking continuum being scaled $ \propto q^2 $ as the one  of the complex energy $ z_\qq $ of the still unobserved continuum mode; although many theoretical works have been devoted to the response functions of a superfluid of fermions, our analytical results on $ \chi $ in this narrow frequency window $ \hbar \omega -2 \Delta = O (q^2) $ are to our knowledge original.  In the BEC-BCS crossover where the order parameter $ \Delta $ is comparable to $ \mu $, the continuum mode causes frequency bumps or dips in the density and modulus response functions of the order parameter. A simple fit of these functions by the sum of a complex resonance and a frequency-affine background makes it possible to estimate the complex energy $ z_\qq $ and the spectral weight of the mode, with a good precision for the modulus response, even when $ \re z_\qq <2 \Delta $ so that the mode is not below the interval of analytic continuation (in the theory) and measurement (in the experiment). This augurs an upcoming observation. 

\section*{Acknowledgments} The interest of this study appeared to us during a discussion with Chris Vale at the BEC 2019 conference in Sant Feliu de Guixols. We also thank Hadrien Kurkjian for helpful remarks on calculating the density-density response function, even if he ultimately preferred to collaborate with other people on the subject \cite{arxiv}. Finally, note that the date of submission of this work is much later than that of the corresponding preprint \cite{hal}; indeed, we had to withdraw our previous submission to a journal which proved unable to produce a referee report.

\end{document}